\documentclass[aps,pra,showpacs,amsmath,amssymb,superscriptaddress,reprint,10pt]{revtex4-2}

\usepackage{graphicx}
\usepackage{dcolumn}
\usepackage{bm}
\usepackage[english]{babel}
\usepackage[utf8]{inputenc}
\usepackage[T1]{fontenc}
\usepackage{epstopdf}
\usepackage{esint}
\usepackage[usenames,dvipsnames]{xcolor}
\usepackage{latexsym}
\usepackage{arydshln}
\usepackage[sans]{dsfont}
\usepackage[colorlinks=true,breaklinks=true,allcolors=blue]{hyperref} % Incompatible with simple use of formulas in headings

\usepackage[title]{appendix} %puts "Appendix" in appendix titles

\usepackage{ifpdf}

\usepackage{babel}
\usepackage{amssymb}
\usepackage{amsfonts}
\usepackage{amsmath}
\usepackage{mathrsfs}
\usepackage{upgreek}  %<-- Use \upmu to avoid capital italicized Greek letters
\usepackage{bbm}
\usepackage{empheq} %to make multiline boxed equations
\usepackage{xspace}
\usepackage{verbatim}
\usepackage{rotating}
\usepackage{afterpage}
\usepackage{sidecap}
\usepackage[super]{nth}
\usepackage{soul}

\DeclareTextSymbol{\degree}{T1}{6}

\newcommand{\e}{\mathrm{e}}
\newcommand{\deriv}{\operatorname{d}\!}

\newcommand{\ket}[1]{\ensuremath{{\left|#1\right\rangle}}\xspace}
\newcommand{\bra}[1]{\ensuremath{{\left\langle #1\right|}}\xspace}

\newcommand{\braOket}[3]{\ensuremath{\langle#1|#2|#3\rangle}\xspace}

\graphicspath{{./Figures/}}

\begin{document}

\title{Mirror-assisted backscattering interferometry to measure the first-order correlation function of the light emitted by quantum scatterers}

\author{P. G. S. Dias}
\affiliation{Departamento de F\'{\i}sica, Universidade Federal de S\~{a}o Carlos, Rod. Washington Lu\'{\i}s, km 235 - SP-310, 13565-905 S\~{a}o Carlos, SP, Brazil}
\author{M. Frometa}
\affiliation{Instituto de F\'{\i}sica de S\~{a}o Carlos, Universidade de S\~{a}o Paulo, C.P. 369, 13560-970 S\~{a}o Carlos, SP, Brazil}
\author{P. H. N. Magnani}
\affiliation{Departamento de F\'{\i}sica, Universidade Federal de S\~{a}o Carlos, Rod. Washington Lu\'{\i}s, km 235 - SP-310, 13565-905 S\~{a}o Carlos, SP, Brazil}
\author{K. R. B. Theophilo}
\affiliation{Instituto de F\'{\i}sica de S\~{a}o Carlos, Universidade de S\~{a}o Paulo, C.P. 369, 13560-970 S\~{a}o Carlos, SP, Brazil}
\affiliation{\textit{Present address:} ICFO – The Institute of Photonic Sciences, Mediterranean Technology Park, Av. Carl Friedrich Gauss, 3, 08860 Castelldefels (Barcelona), Spain}
\author{M. Hugbart}
\affiliation{Universit\'e C\^ote d'Azur, CNRS, INPHYNI, France}
\author{Ph.W. Courteille}
\affiliation{Instituto de F\'{\i}sica de S\~{a}o Carlos, Universidade de S\~{a}o Paulo, C.P. 369, 13560-970 S\~{a}o Carlos, SP, Brazil}
\author{R. Celistrino Teixeira}
\email{teixeira@df.ufscar.br}
\affiliation{Departamento de F\'{\i}sica, Universidade Federal de S\~{a}o Carlos, Rod. Washington Lu\'{\i}s, km 235 - SP-310, 13565-905 S\~{a}o Carlos, SP, Brazil}

\begin{abstract}
We present a new method to obtain the first-order temporal correlation function, $g^{(1)} (\tau)$, of the light scattered by an assembly of point-like quantum scatterers, or equivalently its spectral power distribution. This new method is based on the mirror-assisted backscattering interferometric setup. The contrast of its angular fringes was already linked in the past to the convolution of $g^{(1)} (\tau)$ for different Rabi frequencies taking into account the incoming spatial intensity profile of the probe beam, but we show here that by simply adding a half waveplate to the interferometer in a specific configuration, the fringe contrast becomes $g^{(1)} (\tau)$ of the light scattered by atoms, which are now all subjected to the same laser intensity. This new method has direct application to obtain the saturated spectrum of quantum systems. We discuss some non-trivial aspects of this interferometric setup, and propose an analogy with a double Mach-Zehnder interferometer.
\end{abstract}

\maketitle

%%%%%%%%%%%%%%%%%%%%%%%%%
% Introduction
%%%%%%%%%%%%%%%%%%%%%%%%%

\section{Introduction}

When a two-level quantum system with a non-zero dipolar matrix element is excited by an incoming electromagnetic field, it scatters radiation, and the spectrum of that radiation changes qualitatively between the so-called linear regime, when the average population of the excited level is much smaller than one, to the saturated regime, when the excited population becomes non-negligible and saturates, asymptotically reaching a maximum value of $1/2$. In the linear regime, the scattered light present spectral properties identical to the incoming electromagnetic radiation \cite{CCTAtom}. In the saturated regime, on the other hand, the power spectrum of the scattered light broadens and acquires, for incident monochromatic light, the typical structure of three maxima known as the Mollow triplet \cite{Mollow69}. These maxima can be linked to four possible transitions, two of them of equal frequency, between energy levels in the dressed-state picture of the atom interacting with the incoming electromagnetic field \cite{Schrama92}. This non-linear effect has received recent interest due to the non-classical correlations between photons emitted in different peaks of the spectrum \cite{Schrama92, Peiris15, Carreno17} and the time ordering of photons emitted in different sidebands for non-resonant excitation \cite{Aspect80}, that could be exploited as heralded sources of single photons \cite{Ulhaq12,Portalupi2016} and non-classical light.

The first experimental verification of the saturated spectrum of two-level systems was made with atomic beams \cite{Schuda74,Wu75,Grove77,Walther76}. In those first experiments, the power spectrum of the light scattered by atoms was directly obtained through the use of a Fabry-Perot cavity as a spectral filter, detecting the scattered light power as a function of the frequency.
% (convoluted with the width of the Fabry-Perot transmission spectrum, and eventually modified due to the internal structure of the atomic species). 
Further measurements made with single ions \cite{Stalgies96}, single dye molecules \cite{Wrigge08} and single quantum dots \cite{Flagg09, Peiris15,Lagoudakis2017} applied the same technique to the much fainter signal of those single emitters. For quantum dots \cite{Flagg09}, it was verified that the presence of additional dephasing of the coherences due to the coupling to phonons of the solid state environment make the scattered power spectrum different from the Mollow result.
%In \cite{Wrigge08,Flagg09}, an additional measurement of the second-order temporal correlation function $g^{(2)} (\tau)$, also known as the intensity correlation function, was performed, from which the Rabi frequency of the excitation could be independently measured.

The first-order temporal correlation function $g^{(1)} (\tau)$ is defined as follows:
\begin{equation}\label{eq:g1}
    g^{(1)}(\tau) = \frac{\langle E^\star(t) E(t + \tau) \rangle}{\langle I(t) \rangle},
\end{equation}
with $\langle . \rangle$ corresponding to the averaging over the time $t$, and $I(t) = E^\star(t)E (t)$ the intensity associated to the field $E(t)$. This function is linked to the light spectrum through the Wiener-Khintchine theorem \cite{Loudon}, that states that the power spectrum of the light is proportional to the Fourier transform of $g^{(1)} (\tau)$. This means that the information carried by the power spectrum in frequency domain is equivalent to the information carried by $g^{(1)} (\tau)$ in the temporal domain, and measuring $g^{(1)} (\tau)$ can be considered equivalent to measuring the light power spectrum for a verification of the Mollow theory. The first-order correlation function must be measured through an interferometric measurement, for example with a Michelson interferometer in which the delay $\tau$ is due to the path difference between the two arms, with a self-heterodyne measurement \cite{Okoshi_1980}, or with a heterodyne technique where a laser beam, usually denoted as local oscillator, is superimposed to the light under investigation. This last technique was used to obtain the absolute value of $g^{(1)} (\tau)$ of the light scattered by cold atoms in the linear regime \,\cite{Ferreira_2020}, in which the light is mainly elastically scattered, as well as in the saturated regime, where the light is inelastically scattered, either out of resonance\,\cite{Nakayama10} or at resonance\,\cite{Gutierrez19}.

\begin{figure*} [t]
\centering
\includegraphics[width=0.9\textwidth]{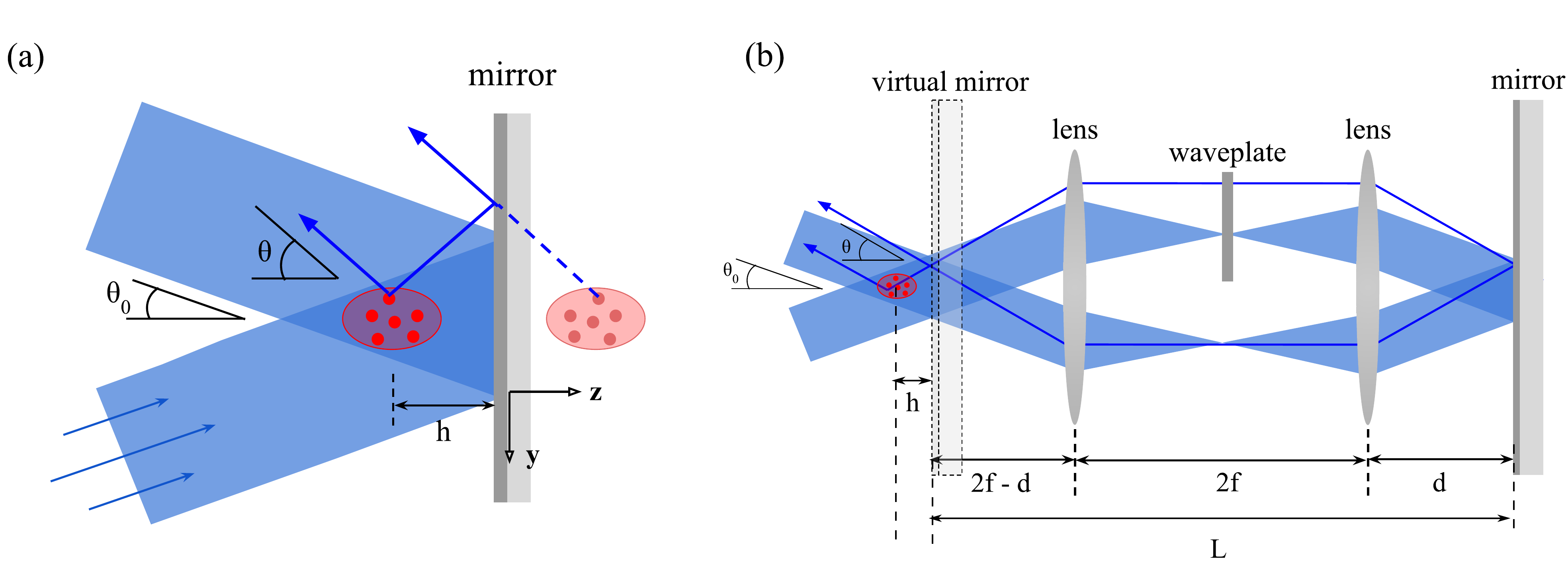}
\caption{(a) Schematic setup of the interferometer. A cloud of atoms (red circles in front of the mirror) is placed at a distance $h$ in front of a mirror. An incoming laser beam (continuous, light blue area and light blue arrows) impinges on the atoms and is reflected by the mirror with a reflection angle $\theta_0$, passing again through the cloud on its way back. The light scattered by the atoms (darker blue arrows), and detected in the far field with an angle $\theta$ with respect to the normal to the mirror, is the superposition of the light directly scattered in this direction, and the light first scattered in the mirror direction and then reflected back to the detector. The scattered light reflected by the mirror can be interpreted as emitted by the image of the atoms (red circles behind the mirror). (b) Physical implementation of the setup. A real mirror is conjugated arbitrarily close to the atoms by a system of converging lenses of equal focal length $f$ and separated by $2f$; in this case, $h$ represents the distance of the virtual mirror to the atoms. This scheme allows one to separate the incoming and reflected light beams between the two lenses, and to insert a half waveplate only on the incoming beam.}
\label{fig:setup}
\end{figure*}

In this article, we report on a new way to obtain $g^{(1)} (\tau)$ for the light scattered by an assembly of quantum scatterers. This method is based on an interferometer called mirror-assisted backscattering (MBS) setup\,\cite{Greffet91, Labeyrie2000observation, Moriya16}. In this scheme, described in more details in the next section and depicted in Fig.\,\ref{fig:setup}a, the scatterers are placed in front of a mirror, such that they are excited by an incident laser and its reflection on the mirror. Accordingly, the scattered light, observed in the far field, is a superposition of the light scattered directly at the observation direction, and the scattered light reflected by the mirror to the observation direction, leading to a fringe pattern in the far field. This scheme relies only on the 
%The specificity of this setup entails some astonishing properties, such as the independence of the fringe contrast on the polarization fluctuations of the reflected light for elastic scattering\,\cite{prep} \MHcomment{a bit annoying to talk about this if not yet published or even submitted ? I would simply remove it}. 
For scattering in the saturated regime, the fringe contrast was shown to be a function of the saturated spectrum emitted by the atoms \cite{Piovella16}. However, in the original setup, this function was a complicated convolution of the saturated spectra emitted by all atoms subject to an intensity spatial modulation caused by the interference of the incoming and reflected excitation beams. In this paper, we show that with a simple polarization rotation of the light that goes from the scatterers to the mirror, we can obtain interference fringes whose contrast is directly the value of the function $g^{(1)} (\tau)$ of the light scattered by the atoms illuminated by twice the incident laser intensity, with $\tau$ the time needed for the light to travel from the scatterers to the mirror and back. 
% \MHcorr{\st{Compared to measurements of $g^{(1)} (\tau)$ based on correlations between time-resolved single photon detection events, this measurement setup does not need fast photodetectors and electronics, being rather comparable, with regard to its implementation, to other interferometric setups such as the Michelson interferometer.}} 
This result has applications in the spectral characterization of any class of identical quantum emitters in the saturated regime, including atoms, ions, molecules and assemblies of identical quantum dots. The spectral characterization allows for example to identify modifications in the electromagnetic modes of the vacuum \cite{Toyli2016resonance,Keitel1995resonance}, to characterize the incident light such as to its intensity and saturation parameter, and to characterize the broadening mechanisms of the transition at work for the scatterers within their environment \cite{Flagg09}. This result also extends the applicability of the MBS setup to the characterization of the coherence of the light scattered by matter; for its previous use in the characterization of the coherence of light scattered by a hot vapour, see \cite{Cherroret2019robust}, where it allowed to identify additional interference rings associated to Raman processes in the multi-level species used for the experiment.

The paper is organized as follows. In Sec.\,\ref{Sec:Setup}, we present the principle of the MBS interferometer in which a half waveplate is added. The total emission profile of a single scatterer in this setup is then calculated in Sec.\,\ref{Sec:SingleScatterer}, and of a spatially extended ensemble of identical scatterers in Sec.\,\ref{Sec:LargeCloud}. We show in this last part that, for a specific position of the half waveplate, one can measure the first-order temporal correlation function of the atoms driven by twice the incident laser intensity. Finally, we conclude on our results in Sec.\,\ref{Sec:Conclusion}.
%discuss our results and propose an analogy with a double Mach-Zehnder interferometer, that allows us to understand some features of this particular interferometric setup.

%%%%%%%%%%%%%%%%%%%%%%%%%
% Setup
%%%%%%%%%%%%%%%%%%%%%%%%%

\section{Mirror-assisted backscattering setup in the presence of a half waveplate} \label{Sec:Setup}

%%%%%%%%%%%%%%%%%%%%%%%%%
% General setup
\subsection{General setup}

The principle of the MBS interferometer is sketched in Fig.\,\ref{fig:setup}a and has already been detailed in Refs.\,\cite{Greffet91, Moriya16}. Briefly, monochromatic coherent light is sent upon the scatterers, with plane wavefront at the scatterers positions, then reflected by a mirror before impinging again on the scatterers. The incident wavevector is defined as
\begin{equation}
 \mathbf{k}_0 = k\left(0,- \sin \theta_0,  \cos \theta_0 \right)
 \end{equation}
with $\theta_0$ the incident angle on the mirror, $k = 2\pi/\lambda$ with $\lambda$ the laser wavelength, and where the z direction determines the normal incidence direction at the mirror. The reflected wavevector corresponds to $\mathbf{k}_0' = k\left(0,- \sin \theta_0, -\cos \theta_0 \right)$. The angular profile of the light scattered from the incident and reflected beams is detected in the far field at an angle $\theta$ with the normal to the mirror and azimuthal angle $\phi$, thus in the direction $\left(\sin \theta \cos \phi, \sin \theta \sin \phi, - \cos \theta \right)$. Due to the presence of the mirror, the scattered light detected in the far field is also the superposition of the light directly emitted in the direction of the detector with wavevector:
\begin{equation}
\mathbf{k} = k\left( \sin \theta \cos \phi,  \sin \theta \sin \phi, -  \cos \theta \right),
\end{equation}
plus the light emitted with wavevector $\mathbf{k}' = k\left( \sin \theta \cos \phi,  \sin \theta \sin \phi, \cos \theta \right)$ and reflected back to the detection direction.

This configuration of light excitation produces angular interference fringes in the scattered light at directions $\theta$ close to $\theta_0$\,\cite{Moriya16,Piovella16}. We consider that $\mathbf{k}_0$, $\mathbf{k}_0'$, $\mathbf{k}$ and $\mathbf{k}'$ are close to the normal direction of the mirror. This particular choice of small incidence angles is always made at experimental implementations of the setup as a compromise between removing the spurious light coming from the incident laser beam on the detector, and the period of the angular fringes, which is inversely proportional to $\theta_0$\,\cite{Moriya16} and must be larger than the detection resolution.

The realized experimental setup is presented in Fig.\,\ref{fig:setup}b. For practical reason, the mirror is placed after two lenses of equal focal length $f$ and separated by $2f$. This creates a virtual image of the real mirror (called virtual mirror) at a distance $2f-d$ from the first lens, with $d$ the distance between the real mirror and the last lens. This particular setup allows placing some optics between the two lenses that acts only once on the incoming light from the scatterers to the mirror. In this paper, we consider the case where we add a half waveplate to control its polarization, as detailed in the next section.

%%%%%%%%%%%%%%%%%%%%%%%%%
% Control of the linear polarization
\subsection{Control of the linear polarization}

The polarization of the incident beam is linear and parallel to the $x$ direction, determined by a unitary vector that we call $\boldsymbol{\epsilon}_x$. We now consider that the polarization of the incident beam after reflection is still linear, but rotated compared to the incident one. This is done by adding a half waveplate after the scattering medium and before the mirror, as shown in Fig.\,\ref{fig:setup}b. The linear polarization after the half waveplate is denoted as  $\boldsymbol{\epsilon}_1$:
\begin{equation}
 \boldsymbol{\epsilon}_1 \, = \, \cos 2 \gamma \, \boldsymbol{\epsilon}_x  \, + \, \sin 2 \gamma \, \boldsymbol{\epsilon}_y.
\end{equation}
with $\gamma$ the angle between the proper axis of the waveplate and $\boldsymbol{\epsilon}_x$.
As said before, we consider that $\theta, \theta_0 \ll 1$ and thus $\boldsymbol{\epsilon}_1.\boldsymbol{\epsilon}_z \simeq 0$. We also write in this limit the action of the waveplate on the polarization, defined by the linear transformation:
\begin{align} 
\mathcal{L}\left[\boldsymbol{\epsilon}_x \right] & = \boldsymbol{\epsilon}_1 \ , \label{eq:L_x}\\
\mathcal{L}\left[\boldsymbol{\epsilon}_1 \right] & = \boldsymbol{\epsilon}_x \ . \label{eq:L_1} 
\end{align}

The total complex electric field seen by a scatterer at position $\mathbf{r} = \left(x,y,z\right)$ (assuming $\mathbf{r} = \left(0,0,0\right)$ at the center of the mirror, as depicted in Fig.\,\ref{fig:setup}a), composed of the incoming plane wave plus the reflected one with rotated polarization, is given by:
\begin{equation}
\mathbf{E}_l(\mathbf{r}) = E_0 \left[\e^{i k \left(\cos \theta_0 z - \sin \theta_0 y\right)} \boldsymbol{\epsilon}_x + \e^{-i k \left(\cos \theta_0 z + \sin \theta_0 y\right)} \boldsymbol{\epsilon}_1 \right] \ .
\label{eq:Elaser}
\end{equation}
We assume that the amplitudes of the incoming and reflected beams are the same. One can note that depending on the waveplate orientation, one goes from interference with full contrast between the incoming and reflected beam when $\boldsymbol{\epsilon}_x.\boldsymbol{\epsilon}_1 = 1$, to no interference when $\boldsymbol{\epsilon}_x.\boldsymbol{\epsilon}_1 = 0$. We can also rewrite this total electric field through its amplitude and direction:
\begin{equation}
\mathbf{E}_l(\mathbf{r}) = E_l(z) \boldsymbol{\epsilon}_l \ ,
\label{eq:El}
\end{equation}
with
\begin{eqnarray}
E_l(z) &=& E_0 \sqrt{2\left[1 + \cos 2 \gamma \, \cos \left(2 k \cos \theta_0 z \right)\right]}, \label{eq:Elmod} \\
 \boldsymbol{\epsilon}_l \, &=& \e^{i k \left(\cos \theta_0 z - \sin \theta_0 y\right)} \frac{  \e^{- 2 i k \cos \theta_0 z}  \boldsymbol{\epsilon}_1  \, +  \, \boldsymbol{\epsilon}_x}{\sqrt{2\left[1 + \cos 2 \gamma \, \cos \left(2 k \cos \theta_0 z \right)\right]}}.  \nonumber\\
 \label{eq:Elpola}
\end{eqnarray}

In what follows, we consider identical quantum point scatterers with a narrow dipolar transition. This is the case, for example, for atoms of the same species with a $J = 0 \rightarrow J = 1$ dipolar transition, from a non-degenerated ground state $\ket{g}$ to an excited state of energy $\hbar \omega_0$ with respect to the ground state. This excited state is composed of three degenerate sublevels $\ket{e_x}$, $\ket{e_y}$ and $\ket{e_z}$, to which the atom can be excited by light linearly polarized respectively in the directions $x$, $y$ and $z$. We call $\Gamma$ the natural width of the transition, originated from the electric dipolar coupling between the atomic transition and the vacuum modes of the quantized electromagnetic radiation, and we consider that the incoming light is resonant: that is, the detuning $\Delta = \omega - \omega_0$ between the frequency of the incoming laser light $\omega = ck$, and the natural frequency $\omega_0$ of the transition, satisfies $\Delta = 0$.
\\

%%%%%%%%%%%%%%%%%%%%%%%%%
% Single scatterer emission profile
%%%%%%%%%%%%%%%%%%%%%%%%%

\section{Single scatterer emission profile} \label{Sec:SingleScatterer}

Let us first calculate the emission profile of only one scatterer at position $\mathbf{r}$. We define the lowering (raising) operators for this atom in the referential rotating with the incoming laser light, $\hat{\sigma}_{\alpha} = \e^{i \omega t} \ket{g}\bra{e_\alpha}$ ($\hat{\sigma}_{\alpha}^{\dagger} = \e^{-i \omega t} \ket{e_\alpha}\bra{g}$), with $\alpha \in \{x,y,z\}$, such that the electric dipole operator $\hat{\mathbf{d}}$ of the atom is given by

\begin{eqnarray}
\hat{\mathbf{d}} &=& d \sum_{\alpha = x,y,z}  \left(\e^{-i \omega t} \hat{\sigma}_{\alpha} + \e^{i \omega t} \hat{\sigma}_{\alpha}^{\dagger}\right) \boldsymbol{\epsilon}_\alpha \\
&=& d \left(\e^{-i \omega t} \hat{\boldsymbol{\sigma}} + \e^{i \omega t} \hat{\boldsymbol{\sigma}}^{\dagger} \right) = \hat{\mathbf{d}}^{(+)} + \hat{\mathbf{d}}^{(-)} \ , 
\label{eq:dipole}
\end{eqnarray}

\noindent with $d$ the amplitude (taken as real without loss of generality, since its phase can be included in the choice of a global phase of each excited level) of the electric dipolar moment of the atomic transition, $d = \braOket{g}{\hat{\mathbf{d}}}{e_\alpha}$ for any $\alpha$; $\boldsymbol{\epsilon}_\alpha$ an unitary vector pointing in the $\alpha$ direction, with $\alpha \in \{x,y,z\}$; the vectorial lowering (raising) operator $\hat{\boldsymbol{\sigma}} = \sum_{\alpha = x,y,z}  \hat{\sigma}_{\alpha} \boldsymbol{\epsilon}_\alpha $ ($\hat{\boldsymbol{\sigma}}^\dagger = \sum_{\alpha = x,y,z}  \hat{\sigma}_{\alpha}^\dagger \boldsymbol{\epsilon}_\alpha $); and the positive and negative frequencies components of the dipole operator, respectively $\hat{\mathbf{d}}^{(+)} = d \, \e^{-i \omega t} \hat{\boldsymbol{\sigma}}$ and $\hat{\mathbf{d}}^{(-)} = d \, \e^{i \omega t} \hat{\boldsymbol{\sigma}}^{\dagger}$. For what follows, we place ourselves in the the Heisenberg picture, with the raising and lowering operators depending on time. 
% and it is instructive to separate their average and quantum fluctuation parts, writing $\hat{\sigma}_{\alpha} (t) = \left<\hat{\sigma}_{\alpha}\right> (t) + \delta \hat{\sigma}_{\alpha} (t)$, with the fluctuating part of the operator, $\delta \hat{\sigma}_{\alpha} (t)$, satisfying $\left<\delta \hat{\sigma}_{\alpha} (t)\right> = 0$. \MHcomment{usefull to say this since we never use $\delta \hat{\sigma}_{\alpha} (t)$ in this paper ?}
We also indicate explicitly the dependence of the atomic coherences on the incoming laser field, $\hat{\sigma}_{\alpha} \equiv \hat{\sigma}_{\alpha} (\mathbf{E}_l(\mathbf{r}), t)$ for an atom at position $\mathbf{r}$. \\

%%%%%%%%%%%%%%%%%%%%%%%%%
% Electric field emitted by a single scatterer
\subsection{Electric field emitted by a single scatterer}

The positive frequency component of the electric field operator of the light scattered by the atom at position $\mathbf{r}$, seen at position $\mathbf{R}$ and time $t$ and emitted with wavevector $\mathbf{k} = k \mathbf{R}/R$ (with $R = |\mathbf{R}|$), before any reflection by the mirror, is expressed in the far-field as\,\cite{Bienaime_2011}:
\begin{eqnarray}
\hat{\mathbf{E}}_{d} (\mathbf{r},\mathbf{R},t) 
&\simeq& \frac{k^2}{4 \pi \epsilon_0 R} \, \hat{\mathbf{d}}^{(+)}\left(\mathbf{E}_l(\mathbf{r}),t_{\text{ret}}\right) \, \e^{i \mathbf{k}(\mathbf{R} - \mathbf{r})} \\
&=& \frac{k^2 d }{4 \pi \epsilon_0 R} \, \hat{\boldsymbol{\sigma}}\left(\mathbf{E}_l(\mathbf{r}),t_{\text{ret}}\right) \, \e^{i \mathbf{k}(\mathbf{R} - \mathbf{r})} \, \e^{-i \omega t}, \nonumber \\
\label{eq:Ed}
\end{eqnarray}
with $\epsilon_0$ the vacuum permittivity, $t_{\text{ret}} (\mathbf{R},\mathbf{r},t) = t - \frac{\mathbf{k}(\mathbf{R} - \mathbf{r})}{kc}$ the instant at which the light was emitted to be detected at time $t$ in position $\mathbf{R}$, $c$ the vacuum speed of light, and where we have used the approximation $\theta, \theta_0 \ll 1$ implying $\hat{\mathbf{d}}^{(+)}.\mathbf{k} \simeq 0$.

The mirror also reflects the light emitted in direction $\mathbf{k}'$ back to the direction $\mathbf{k}$. Before reflection, the electric field passes through the waveplate, suffering the linear transformation $\mathcal{L}$ which acts on $\hat{\boldsymbol{\sigma}}\left(\mathbf{E}_l(\mathbf{r}),t_{\text{ret}}\right)$. The reflected scattered electric field detected at point $R$ can be written as follows:
\begin{equation}
\hat{\mathbf{E}}_{r} (\mathbf{r},\mathbf{R},t)  \simeq \frac{k^2 d }{4 \pi \epsilon_0 R} \, \mathcal{L} \left[\hat{\boldsymbol{\sigma}}\left(\mathbf{E}_l(\mathbf{r}),t_{\text{ret}}'\right)\right] \, \e^{i \mathbf{k}(\mathbf{R} - \mathbf{r}')} \, \e^{-i \omega t} \ ,
\end{equation}
where $\mathbf{r}' = (x,y,-z)$ is the position of the mirror image of the atom at position $\mathbf{r}$, and  $t_{\text{ret}}' \equiv t_{\text{ret}}  (\mathbf{R},\mathbf{r}',t) = t - \frac{\mathbf{k}(\mathbf{R} - \mathbf{r}')}{kc} -\frac{2L}{c}$ the retarded time for the reflected emission of the atom, which also depends on $L$, the path length between the virtual and real mirrors. We see that this reflected electric field depends on the electric field at position $\mathbf{r}$, as the scattered field is emitted by the atom at position $\mathbf{r}$, but it has a different spatial phase and time delay with respect to the electric field directly emitted in the wavevector $\mathbf{k}$.

The total scattered electric field emitted by one atom at $\mathbf{r}$ and detected at position $\mathbf{R}$ and time $t$ is the sum of both components:
\begin{eqnarray}
\hat{\mathbf{E}}_{1} (\mathbf{r},\mathbf{R},t) &=&  \hat{\mathbf{E}}_{d} (\mathbf{r},\mathbf{R},t) + \hat{\mathbf{E}}_{r} (\mathbf{r},\mathbf{R},t) \\
&=& \frac{k^2 d }{4 \pi \epsilon_0 R} \, \e^{i \mathbf{k}(\mathbf{R} - \mathbf{r})} \, \e^{-i \omega t} \, \Bigg[\hat{\boldsymbol{\sigma}}\left(\mathbf{E}_l(\mathbf{r}),t_{\text{ret}}\right) \, \nonumber\\
 && +  \, \e^{2 i k \cos \theta z} \, \mathcal{L}\left[\hat{\boldsymbol{\sigma}}\left(\mathbf{E}_l(\mathbf{r}),t_{\text{ret}}'\right)\right] \Bigg] \ ,
\end{eqnarray}
The direction of the atomic dipole operator is in the same direction as the incoming electric field seen by the atom \cite{CCTAtom}, such that we can write $\hat{\boldsymbol{\sigma}}\left(\mathbf{E}_l(\mathbf{r}),t_{\text{ret}}\right) = \hat{\sigma}\left(E_l(z), t_{\text{ret}}\right) \, \boldsymbol{\epsilon}_l (\mathbf{r})$, where the scalar operator $ \hat{\sigma} (E,t)$ represents the rising operator of a two-level system at time $t$, subject to a scalar electric field excitation of modulus $E$. We can thus write:
\begin{multline}
\hat{\mathbf{E}}_{1} (\mathbf{r},\mathbf{R},t) = \frac{k^2 d }{4 \pi \epsilon_0 R} \e^{i \mathbf{k}(\mathbf{R} - \mathbf{r})} \, \e^{-i \omega t} \Bigg[\hat{\sigma}\left(E_l(z),t_{\text{ret}}\right) \boldsymbol{\epsilon}_l (\mathbf{r}) \\
 +  \, \e^{2 i k \cos \theta z} \, \hat{\sigma}\left(E_l(z),t_{\text{ret}}'\right) \mathcal{L}\left[\boldsymbol{\epsilon}_l (\mathbf{r})\right] \Bigg] \ .
\label{eq:E1}
\end{multline}

%%%%%%%%%%%%%%%%%%%%%%%%%
% Intensity emitted by a single scatterer
\subsection{Intensity emitted by a single scatterer}

The intensity emitted by this atom, and detected at position $R$ and time $t$, is given by
\begin{eqnarray}
&& I_1 (\mathbf{r},\mathbf{R},t) = \frac{\epsilon_0 c}{2} \left<\hat{\mathbf{E}}_{1}^{\dagger} (\mathbf{r},\mathbf{R},t) \, \hat{\mathbf{E}}_{1} (\mathbf{r},\mathbf{R},t) \right>  \\
&=& \frac{k^4 d^2 c}{32 \pi^2 \epsilon_0 R^2} \Big\{ \left< \hat{\sigma}^{\dagger}\left(E_l(z),t_{\text{ret}} \right) \hat{\sigma}\left(E_l(z),t_{\text{ret}} \right) \right> \boldsymbol{\epsilon}_l ^{\dagger}(\mathbf{r}).\boldsymbol{\epsilon}_l (\mathbf{r}) \nonumber \\
&& + \left< \hat{\sigma}^{\dagger}\left(E_l(z),t_{\text{ret}}' \right) \hat{\sigma}\left(E_l(z),t_{\text{ret}}'\right) \right> \mathcal{L}\left[\boldsymbol{\epsilon}_l ^{\dagger} (\mathbf{r})\right].\mathcal{L}\left[\boldsymbol{\epsilon}_l (\mathbf{r})\right]  \nonumber \\
&& + \e^{2 i k \cos \theta z}  \left< \hat{\sigma}^{\dagger}\left(E_l(z),t_{\text{ret}} \right) \hat{\sigma}\left(E_l(z),t_{\text{ret}}' \right) \right> \boldsymbol{\epsilon}_l ^{\dagger} (\mathbf{r}).\mathcal{L}\left[\boldsymbol{\epsilon}_l (\mathbf{r})\right]   \nonumber \\
&& +\e^{- 2 i k \cos \theta z}  \left< \hat{\sigma}^{\dagger}\left(E_l(z),t_{\text{ret}}'\right) \hat{\sigma}\left(E_l(z),t_{\text{ret}}\right) \right> \mathcal{L}\left[\boldsymbol{\epsilon}_l ^{\dagger} (\mathbf{r})\right].\boldsymbol{\epsilon}_l (\mathbf{r})  \Big\}. \nonumber\\
&&
\end{eqnarray}
According to Eqs.\,(\ref{eq:L_x}), (\ref{eq:L_1}) and (\ref{eq:Elpola}), the polarization parts become:
\begin{align}
 \mathcal{L}\left[\boldsymbol{\epsilon}_l (\mathbf{r})\right] & \nonumber\\
 = \e^{i k \left(\cos \theta_0 z - \sin \theta_0 y\right)} & \frac{  \e^{- 2 i k \cos \theta_0 z}  \boldsymbol{\epsilon}_x  \, +  \, \boldsymbol{\epsilon}_1}{\sqrt{2\left[1 + 1 \cos 2 \gamma \, \cos \left(2 k \cos \theta_0 z \right)\right]}} \ .\\
\boldsymbol{\epsilon}_l ^{\dagger}(\mathbf{r}).\boldsymbol{\epsilon}_l (\mathbf{r}) & = 1  \ ,\\
\mathcal{L}\left[\boldsymbol{\epsilon}_l  ^{\dagger} (\mathbf{r})\right] . \mathcal{L}\left[\boldsymbol{\epsilon}_l (\mathbf{r})\right]  & = 1 \ , \\
\boldsymbol{\epsilon}_l ^{\dagger}(\mathbf{r}).\mathcal{L}\left[\boldsymbol{\epsilon}_l (\mathbf{r})\right] & = \mathcal{L}\left[\boldsymbol{\epsilon}_l  ^{\dagger} (\mathbf{r})\right] . \boldsymbol{\epsilon}_l (\mathbf{r}) \nonumber \\
 & = \frac{\cos 2 \gamma \, + \, \cos \left(2 k \cos \theta_0 z \right)}{1 \, + \, \cos 2 \gamma \, \cos \left(2 k \cos \theta_0 z \right)} \ .
\end{align}

We are interested in the intensity at a time $t \rightarrow \infty$, that is, the steady state configuration, after all transients of the atomic response to the incoming electric field have decayed to zero. In this steady state regime, the average values of the product of coherences only depend on the relative time between those coherences. For an excitation at resonance, this can be expressed as \cite{Piovella16}
\begin{multline}
\left< \hat{\sigma}^{\dagger}\left(E_l(z),t_{\text{ret}}\right) \hat{\sigma}\left(E_l(z),t_{\text{ret}}\right) \right> \\
= \left< \hat{\sigma}^{\dagger}\left(E_l(z),t_{\text{ret}}'\right) \hat{\sigma}\left(E_l(z),t_{\text{ret}}'\right) \right> = \frac{s(z)}{2 (1 + s(z))} \ ,
\end{multline}
\begin{multline}
\left< \hat{\sigma}^{\dagger}\left(E_l(z),t_{\text{ret}}'\right) \hat{\sigma}\left(E_l(z),t_{\text{ret}}\right) \right> \\
= \left< \hat{\sigma}^{\dagger}\left(E_l(z),t_{\text{ret}}\right) \hat{\sigma}\left(E_l(z),t_{\text{ret}}'\right) \right> = \frac{s(z)}{2 (1 + s(z))} \, \tilde{g}^{(1)}_{z} (\tau_c) \ ,
\end{multline}
with $s(z)$ the saturation parameter at position $z$. The quantity $\tilde{g}^{(1)}_{z} (\tau_c)$ corresponds to the first-order temporal correlation function of the electric field emitted by the atom at position $z$ from the virtual mirror, in the referential frame rotating with the frequency of the incoming laser light, as a function of $\tau_c =  t_{\text{ret}} - t_{\text{ret}}' = \mathbf{k}.(\mathbf{r}-\mathbf{r}')/kc +2L/c = 2 z \cos \theta/c +2L/c$. 
%For a virtual mirror placed at an average distance $h$ from the scatterers much bigger than the size of the cloud of scatterers, and angles $\theta \simeq \theta_0$, one can write $\tau_c \simeq 2 h \cos \theta_0 / c +2L/c$, a parameter independent of $z$. The first-order correlation function in the rotating frame of the laser light is given by \cite{Mollow69, Piovella16}
\begin{widetext}
\begin{equation}
 \tilde{g}^{(1)}_{z} (\tau_c) = \frac{1}{1 + s(z)} + \frac{1}{2} \left[\e^{-\Gamma \tau_c/2}  + \frac{s(z) - 1}{s(z) + 1} \cos \left(\Omega_M(z) \tau_c \right) \e^{-3 \Gamma \tau_c/4} + \frac{\Gamma}{4\Omega_M(z)} \frac{5s(z) - 1}{s(z) + 1} \sin \left(\Omega_M(z) \tau_c \right) \e^{-3 \Gamma \tau_c/4} \right] \ ,
 \label{eq:g1_z}
\end{equation}
\end{widetext}
where $\Omega_l(z) = d  E_l(z)/\hbar = \Gamma \sqrt{s(z)/2}$ is the scalar, real Rabi frequency, and $\Omega_M (z) = \sqrt{\Omega_l^2(z) - \Gamma^2/16}$. The first term in the RHS of eq.~(\ref{eq:g1_z}) is independent of $\tau$ and represents the correlations on the light coherently scattered \cite{CCTAtom}, that have same spectrum as the incoming monochromatic light, while the other terms correspond to the correlations for the light incoherently scattered, that present in the frequency domain the typical broadened structure of three peaks known as the Mollow triplet. The total intensity scattered by the atom in steady state is finally given by
\begin{multline}
I_1 (\mathbf{k},\mathbf{r}) = I_a \frac{s(z)}{1 + s(z)} \Bigg[1 + \\
\tilde{g}^{(1)}_z (\tau_c) \, \frac{\cos 2 \gamma \, + \, \cos \left(2 k z \cos \theta_0 \right)}{1 \, + \, \cos 2 \gamma \, \cos \left(2 k z \cos \theta_0 \right)} \cos \left(2 k z \cos \theta \right) \Bigg] \ .
\label{eq:I1}
\end{multline}
with $I_a \equiv \frac{k^4 d^2 c}{32 \pi^2 \epsilon_0 R^2}$.

%%%%%%%%%%%%%%%%%%%%%%%%%
% Scattered intensity for parallel polarization
\subsection{Scattered intensity for parallel polarization}

\begin{figure} [t]
\centering
\includegraphics[width=\columnwidth]{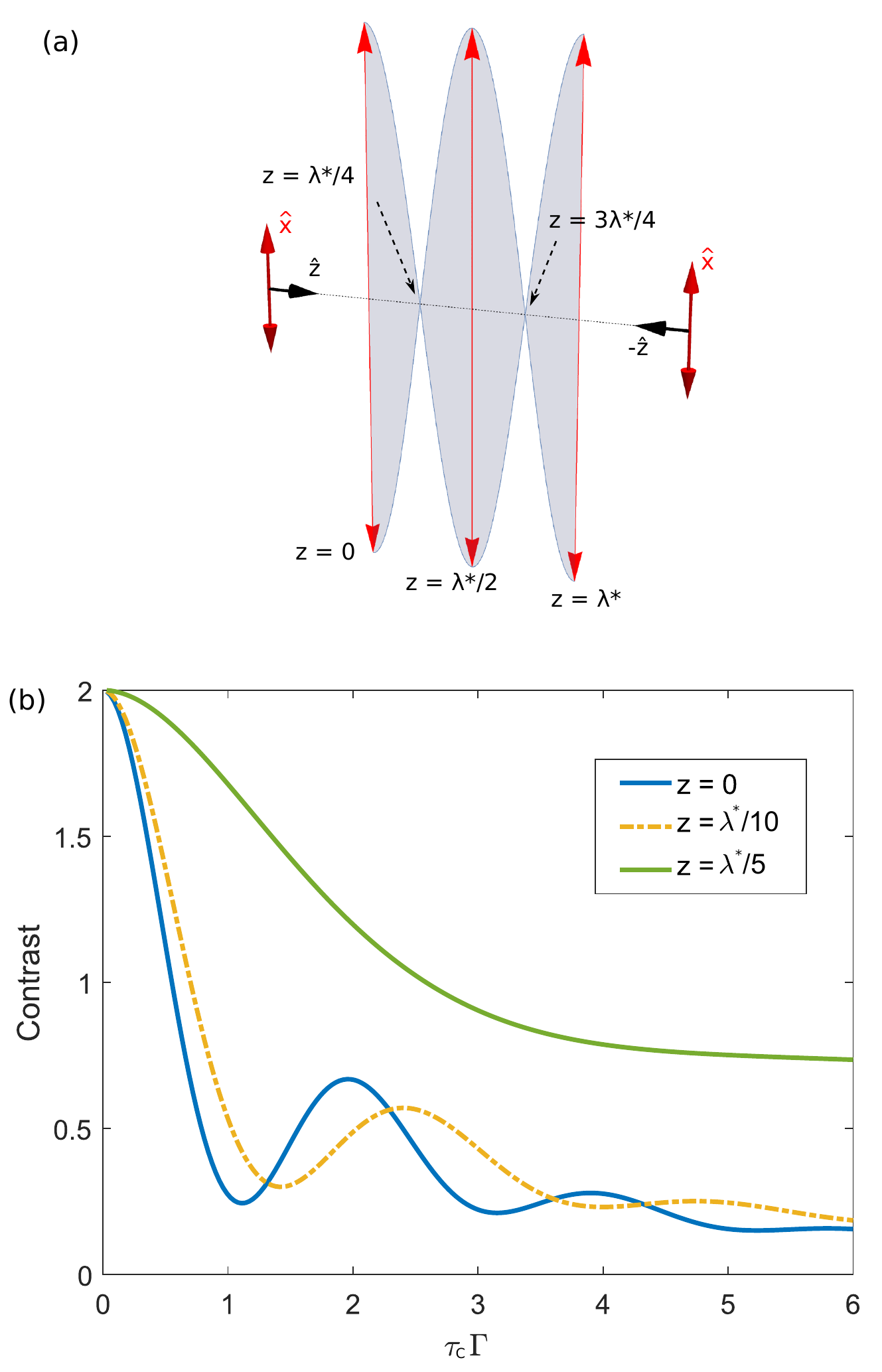} \\
\caption{(a) Amplitude modulation of the total laser electric field when the polarization of the reflected beam is parallel to the incident one ($\gamma = 0$). The incoming and reflected beams with same polarization create an intensity grating in space with a spatial period $\lambda^*/2 = \lambda/(2\cos(\theta_0))$. (b) Contrast of the fringes of the light scattered by a single atom, upon incidence of a plane wave with saturation parameter $s_0 = 5$, as a function of $\tau_c$ for the $\gamma = 0$ case, Eq.\,(\ref{eq:C1parallel}), and for different atomic positions.}
\label{fig:fringes_parallel_polar}
\end{figure}

Let's take a look at this last expression in two extreme cases, when the polarization of the light that went to the mirror and back is parallel to the incident polarization or perpendicular. In the first one, the angle $\gamma$ of the proper axis of the waveplate with the direction $\boldsymbol{\epsilon}_x$ is $\gamma = 0$. In this situation, the waveplate does not affect the light polarization, that remains fully linear and parallel to the $\boldsymbol{\epsilon}_x$ direction. The incoming and reflected laser beams with same polarization create an intensity grating in space along $z$: $I_l(z) = 4 E_0^2 \cos^2(kz\cos\theta_0)$, as represented in Fig.\,\ref{fig:fringes_parallel_polar}a. This modulates the Rabi frequency seen by the atoms $\Omega_l (z)$, the frequency $\Omega_M(z)$, as well as the saturation parameter $s(z)$.

We call the intensity scattered by one atom at position $z$ in this configuration $I_{1,\parallel}$\,\cite{Piovella16}:
\begin{equation}
I_{1,\parallel} (\mathbf{k},\mathbf{r}) = I_a \frac{s(z)}{1 + s(z)} \left[1 + \tilde{g}^{(1)}_z (\tau_c) \, \cos \left(2 k z \cos \theta \right) \right] \ .
\label{eq:I1parallel}
\end{equation}
The term $\cos \left(2 k z \cos \theta \right)$ comes from the interference between the scattered light sent directly to the detector, and the scattered light reflected by the mirror, with $2 k z \cos \theta$ the phase difference between both paths. This leads to an angular interference pattern forming fringes with an angular period $\pi/kz \theta_0$ for $\theta \simeq \theta_0 \ll 1$. The contrast of the fringes, defined as the amplitude peak-to-peak of the fringes divided by the mean intensity, is given by:
\begin{equation}
C_{1,\parallel} = 2 \left|\tilde{g}^{(1)}_z (\tau_c) \right|.
\label{eq:C1parallel}
\end{equation}
It depends on the delay $\tau_c$ and on the position of the atom $z$.

\begin{figure} [t]
\centering
\includegraphics[width=\columnwidth]{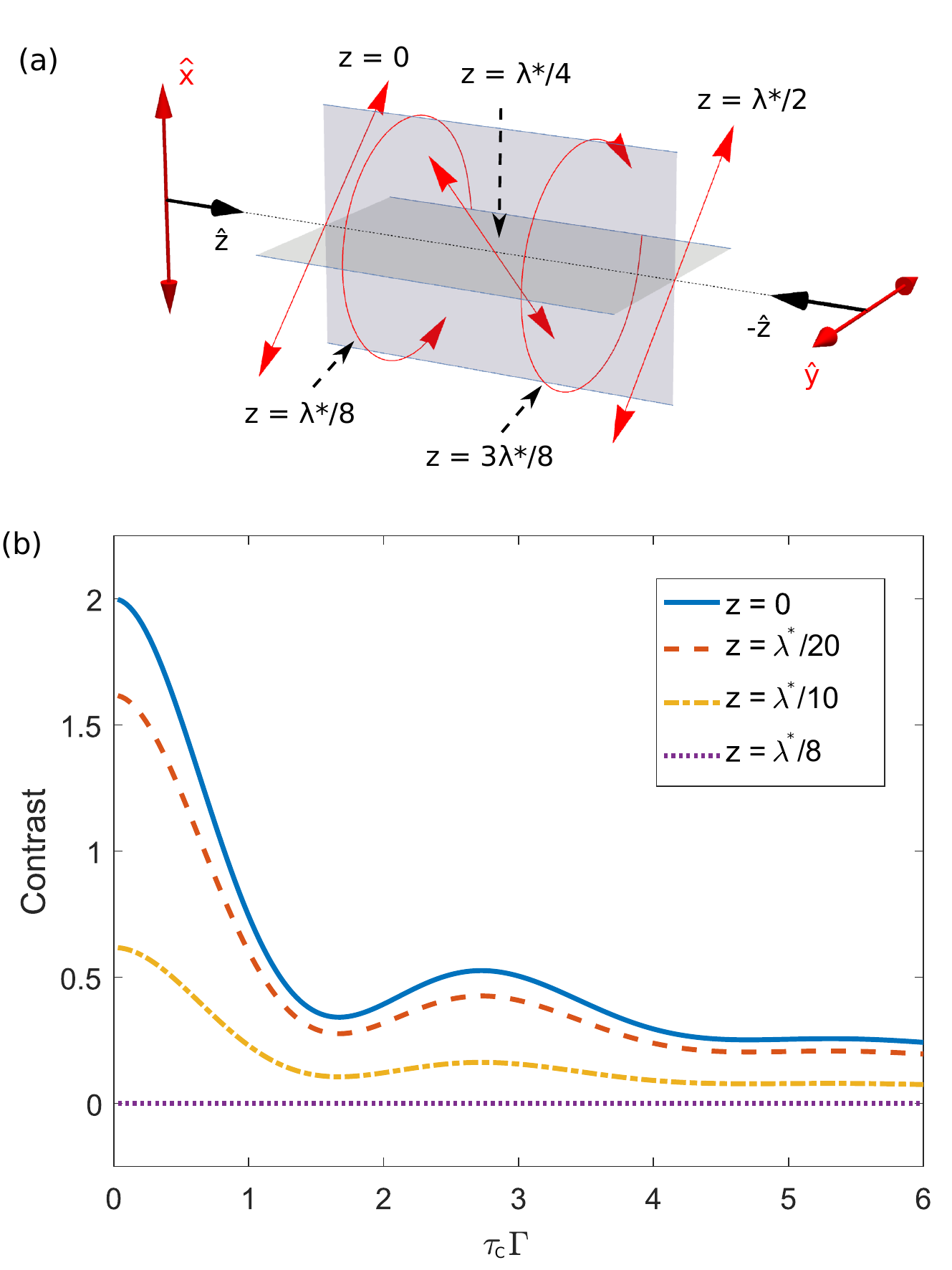} \\
\caption{(a) Polarization modulation of the total laser electric field when the polarization of the reflected beam is perpendicular to the incident one ($\gamma = \pi/4$). The incoming and reflected beams with orthogonal polarization create a polarization grating in space with a spatial period $\lambda^*/2 = \lambda/(2\cos(\theta_0))$. (b) Contrast of the fringes of the light scattered by a single atom, upon incidence of a plane wave with saturation parameter $s_0 = 5$, as a function of $\tau_c$ for the  $\gamma = \pi/4$ case, Eq.\,(\ref{eq:C1perp}), and for different atomic positions.}
\label{fig:fringes_perp_polar}
\end{figure}

This contrast is plotted in Fig.\,\ref{fig:fringes_parallel_polar}b as a function of $\tau_c$, with $\theta \simeq \theta_0 = 4.3^{\circ}$ for different positions $z$. The incoming plane wave has an intensity that corresponds to a saturation parameter $s_0 = 2 \left(\tfrac{d E_0}{\hbar \Gamma}\right)^2 = 5$. Each one of those curves correspond to twice the value of $|\tilde{g}^{(1)}_z (\tau_c) |$ for the light scattered by the atom at its position $z$. Their shape differ qualitatively, as expected, because the Rabi frequency varies with the $z$ position. The $z=0$ case corresponds to the blue line in Fig.\,\ref{fig:fringes_parallel_polar}b. The incoming and reflected laser beams fully interfere, leading to a saturation parameter $4s_0 = 20$ and thus a Rabi frequency  $\Omega_l /\Gamma \simeq 3.16$ and $\Omega_M/\Gamma \simeq 3.16$. The two sidebands of the Mollow triplet emitted by the atoms beat with its carrier, leading to a pseudo-period in the time domain of $\tau_c \Gamma = 2\pi \Gamma/\Omega_M \simeq 2$ (see Eq.\,(\ref{eq:g1_z})), as observed on the blue line. When $z$ is increased within the first half-period of the grating, the intensity decreases, as well as the Rabi frequency. The pseudo-period increases accordingly (yellow dashed line in Fig.\,\ref{fig:fringes_parallel_polar}b). Finally, when the intensity is close to zero, no beating is observed anymore because the light is scattered mostly elastically. The small inelastic component of the spectrum is also less broadened, losing its characteristic shape with three maxima for $\Omega_l \lesssim \Gamma/2$. The temporal decay is dominated by the terms $\e^{-\Gamma \tau_c/2}$ and $\e^{-3 \Gamma \tau_c/4}$ as can be seen in Eq.\,(\ref{eq:g1_z}) (green line in Fig.\,\ref{fig:fringes_parallel_polar}b).

%%%%%%%%%%%%%%%%%%%%%%%%%
% Scattered intensity for perpendicular polarization

\subsection{Scattered intensity for perpendicular polarization}

The opposite situation is for $\gamma = \pi/4$. In this case, the polarization direction $\boldsymbol{\epsilon}_x$ is rotated by $90^\circ$ into $\boldsymbol{\epsilon}_y$, and vice-versa. The incident and reflected beams do not interfere anymore, leading to a constant total laser electric field along $z$: $E_l(z) = \sqrt{2} E_0$. However, the orthogonal polarizations create a polarization grating, with same spatial periodicity as the intensity grating for the case $\gamma = 0$. As shown in Fig\,\ref{fig:fringes_perp_polar}a,  the polarization of the total light seen by an atom varies from linear at the direction $(\boldsymbol{\epsilon}_x + \boldsymbol{\epsilon}_y)/\sqrt{2}$ (that is, aligned with the proper axis of the waveplate), to circular, to linear at the orthogonal direction (thus aligned with the second axis of the waveplate), and back, when the position $z$ is scanned within one grating period.

Accordingly, all parameters that depend on $z$ through $E_l(z)$ become constant: $s(z) \equiv s = 2 s_0$, $\Omega_l(z) \equiv \Omega_l$, $\Omega_M (z) \equiv \Omega_M$, and ultimately $\tilde{g}^{(1)}_z (\tau_c) \equiv \tilde{g}^{(1)} (\tau_c)$. In this situation, the total intensity scattered by the atom $I_{1, \perp}$ is
\begin{multline}
I_{1,\perp} (\mathbf{k},\mathbf{r}) = I_a \frac{s}{1 + s} \Bigg[1 + \\
\tilde{g}^{(1)} (\tau_c) \, \cos \left(2 k z \cos \theta_0 \right) \, \cos \left(2 k z \cos \theta \right) \Bigg] \ .
\label{eq:I1perp}
\end{multline}
We still have an angular interference pattern with a contrast that is given by:
\begin{equation}
C_{1,\perp} = 2\left|\tilde{g}^{(1)}(\tau_c) \cos(2kz\cos\theta_0) \right|.
\label{eq:C1perp}
\end{equation}

This contrast is plotted for different atomic positions $z$ in Fig.\,\ref{fig:fringes_perp_polar}b as a function of $\tau_c$, for same conditions as for Fig.\,\ref{fig:fringes_parallel_polar}b: $s_0 = 5$ and $\theta \simeq \theta_0 = 4.3^{\circ}$. It is clear that all curves are identical up to a pre-factor, the $\cos \left(2 k z \cos \theta_0 \right)$ term, as expected from Eq.\,(\ref{eq:C1perp}). For $z = 0$, the laser beam polarization is linear, parallel to one of the axis of the half waveplate, as shown in  Fig\,\ref{fig:fringes_perp_polar}a. Since the light scattered by an atom at $\theta \sim \theta_0$ has same polarization as the light seen by it, the light scattered directly to the detector and the light scattered and reflected have the same linear polarization. They will thus fully interfere on the detector, corresponding to maximum contrast (blue line in  Fig.\,\ref{fig:fringes_perp_polar}b. On the contrary, for $z = \lambda^*/8$, the polarization of the total light seen by the atom is circular. The light directly scattered to the detector keeps this same circular polarization, while the scattered light reflected by the mirror is orthogonally circularly polarized after passage trough the half waveplate. This leads to no interference on the detector and thus null contrast (purple dotted line in  Fig.\,\ref{fig:fringes_perp_polar}b).

Finally, in the general case corresponding to any value of $\gamma$, both the amplitude and polarization of the electric field that excites the atoms are periodically modulated, with the same spatial period as for the particular cases discussed above, and no simple interpretation is possible for the contrast of the single atom.

%%%%%%%%%%%%%%%%%%%%%%%%%
%Discussion of the single-atom case
%%%%%%%%%%%%%%%%%%%%%%%%%

\subsection{Discussion of the single-atom case}

\begin{figure*} [t]
\centering
\includegraphics[width=1.8\columnwidth]{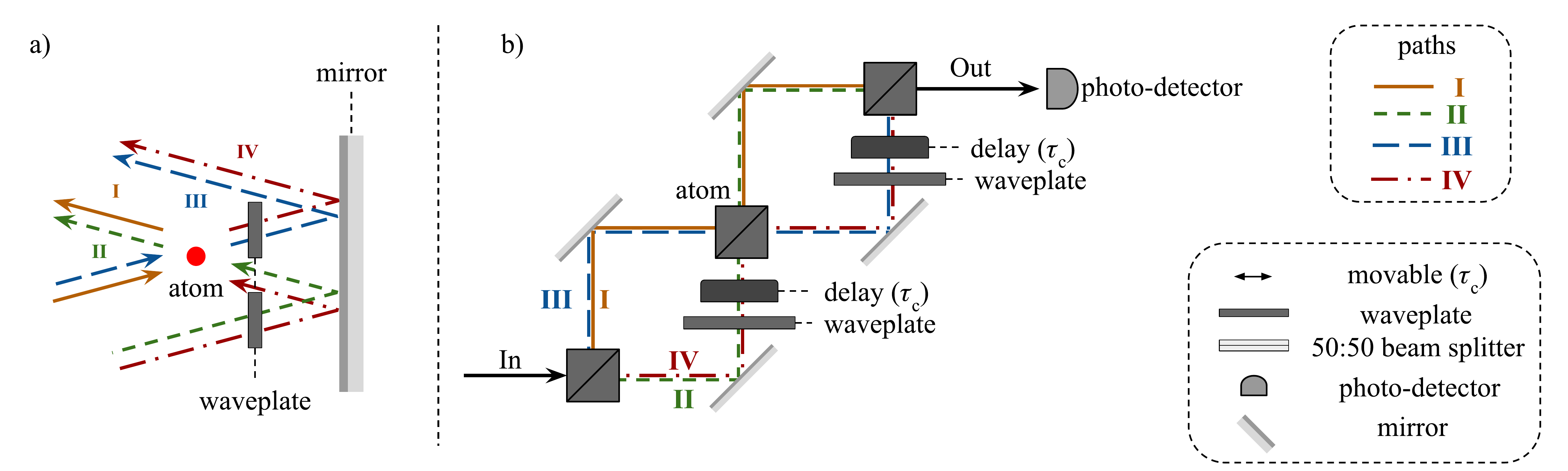}
\caption{(a) In the linear regime, the interference fringes for a single atom are the result of the coherent superposition of four different paths for the scattered photons. They are identified by different colors and traces (continous, short dashed, dashed, dash-dotted) in the drawing. (b) The scheme shown in (a) is formally equivalent to a double Mach-Zehnder interferometer, where the atom is represented by a beamsplitter in the regime with $s_0 \ll 1$. The reflection by the mirror in presence of the waveplate can be represented by adding a time delay and a polarization rotation to both lower arms, in a correlated way. When the two-level scatterer becomes saturated, this comparison fails. Nevertheless, by replacing the central cube by the saturated atom, only source of non-linearities of the system, the analogy becomes again accurate.}
\label{fig:Mach-Zehnder}
\end{figure*}

As other more common interferometer setups, the MBS effect relies on the interference of light that went through at least two different paths from the same source to the same detection event. For a single atom, the light detected at the far field is the superposition of the light scattered by it, and sent to two different directions: either to the detector, or first to the mirror and then reflected back to the detector. But the light that excites the atom is already a superposition of two different paths: the light either impinges on the atom directly, or after reflection by the mirror. The MBS effect relies on both interferences, which allows for the survival of the interference fringes for all linear polarization rotations, even when the rotation angle is equal to $\pi/2$. This double interference implies that we have in total four different amplitudes, associated to four different paths, that add up coherently to form the total amplitude of the electric field of the scattered light at the detector. These paths are shown in Fig. \ref{fig:Mach-Zehnder}a. Now, we see that path I contains no reflection, and so it doesn't pass through the waveplate, while path IV crosses it twice: both paths thus have the same polarization. On the other hand, paths II and III contain only one reflection, and they have the same rotated polarization. We will thus always have the paths interfering at least two by two at the detector, preserving always some interference effect for any polarization rotation.

The interpretation in terms of a double interference, one for the incoming light and one for the scattered light, brings another interesting image to the MBS effect: for a single atom in the linear regime, this interferometric setup is equivalent to a double Mach-Zehnder (MZ) interferometer, with correlated delays and polarization shifts on both lower arms, where the central non-polarizing beamsplitter stands for the atom itself. We represent this equivalent system in Fig. \ref{fig:Mach-Zehnder}b.
For scattering in the linear regime, the behaviour of the atom is equivalent to a non-polarizing beamsplitter up to numerical factors that depend only on its scattering differential cross-section. For the saturated regime, on the other hand, no simple linear device can mimic the behaviour of the scatterers; its response will be a non-linear function of the total input. Replacing the central cube by a scatterer with its specific non-linearity keeps the double MZ interferometer still an accurate picture for the single scatterer behaviour.

%%%%%%%%%%%%%%%%%%%%%%%%%
%Large, disordered cloud of scatterers
%%%%%%%%%%%%%%%%%%%%%%%%%

\section{Extended clouds of ramdomly distributed scatterers} \label{Sec:LargeCloud}

We now consider the problem of the fluorescence profile of a cloud of scatterers, of dimensions much bigger than the wavelength of light. Specifically, we suppose an atomic cloud of $N$ atoms, with an average Gaussian density profile $\rho(\mathbf{r})$ given by
\begin{equation}
\rho(\mathbf{r}) = \frac{N}{(2\pi)^{3/2} s_z s_r^2} \e^{-\frac{(x^2 +y^2)}{2 s_r^2} - \frac{(z+h)^2}{2 s_z^2}} \ ,
\label{eq:density}
\end{equation}
\noindent where $h$ is the distance between the center of the atomic cloud and the virtual mirror, $s_r$ its transverse size, and $s_z$ its longitudinal size. In order to calculate the total light intensity scattered by the atomic cloud, we would need to consider the total electric field emitted by all individual scatterers. But, following \cite{Piovella16}, the averaging over all atomic positions for a cloud with transverse and longitudinal sizes $s_r, s_z \gg \lambda$ makes the interference between the light scattered by different atoms average out to zero. We thus end up with the total intensity being equal to the incoherent sum of the intensities emitted by each atom, that can be written in the limit of large N as
%\begin{equation}
%I(\mathbf{k}) = \int_{\mathbb{R}^3} \deriv^3 r \, \rho(\mathbf{r}) \, I_1(\mathbf{k}, \mathbf{r}) \ .
%\label{eq:Int}
%\end{equation}
%
\begin{multline}
I(\mathbf{k})  = \int_{\mathbb{R}^3} \deriv^3 r \, \rho(\mathbf{r}) \, I_1(\mathbf{k}, \mathbf{r}) =\\
\frac{N I_a}{\sqrt{2\pi} s_z} \int \deriv z \, \e^{- \frac{(z+h)^2}{ 2 s_z^2}}  \, \frac{s(z)}{1 + s(z)} \\
\left[1 + \tilde{g}^{(1)}_z (\tau_c) \, \frac{\cos 2 \gamma \, + \, \cos \left(2 k z \cos \theta_0 \right)}{1 \, + \, \cos 2 \gamma \, \cos \left(2 k z \cos \theta_0 \right)} \cos \left(2 k z \cos \theta \right) \right] .\\
\label{eq:Int_I}
\end{multline}
Note that $\tau_c$ is considered to be independent on $z$. This is justified for $s_z \ll L$, in which case the distance between the real mirror and each atom is almost the same and equal to $L$. It is important to note that this averaging suppose that the density of the atomic cloud satisfies $\rho(\mathbf{r}) \ll k^3$, and the optical density $b_0$ in the $z$ direction satisfies $b_0 \ll 1$. Indeed, on one hand, keeping the optical density low is necessary for neglecting the attenuation of the incoming laser light across the cloud, ensuring that all atoms see an incoming light with same electric field $E_0$. On the other hand, keeping the density and optical density low allow us to neglect any collective effects on the light emission by the atomic cloud, which is important since the MBS effect is a single-atom effect. This sets a limit on the number of atoms that the experimentalist can afford for a specific atomic geometry, stated above as a function of the density and optical density of the atomic cloud.

\begin{figure}[t]
\centering
\includegraphics[width=\columnwidth]{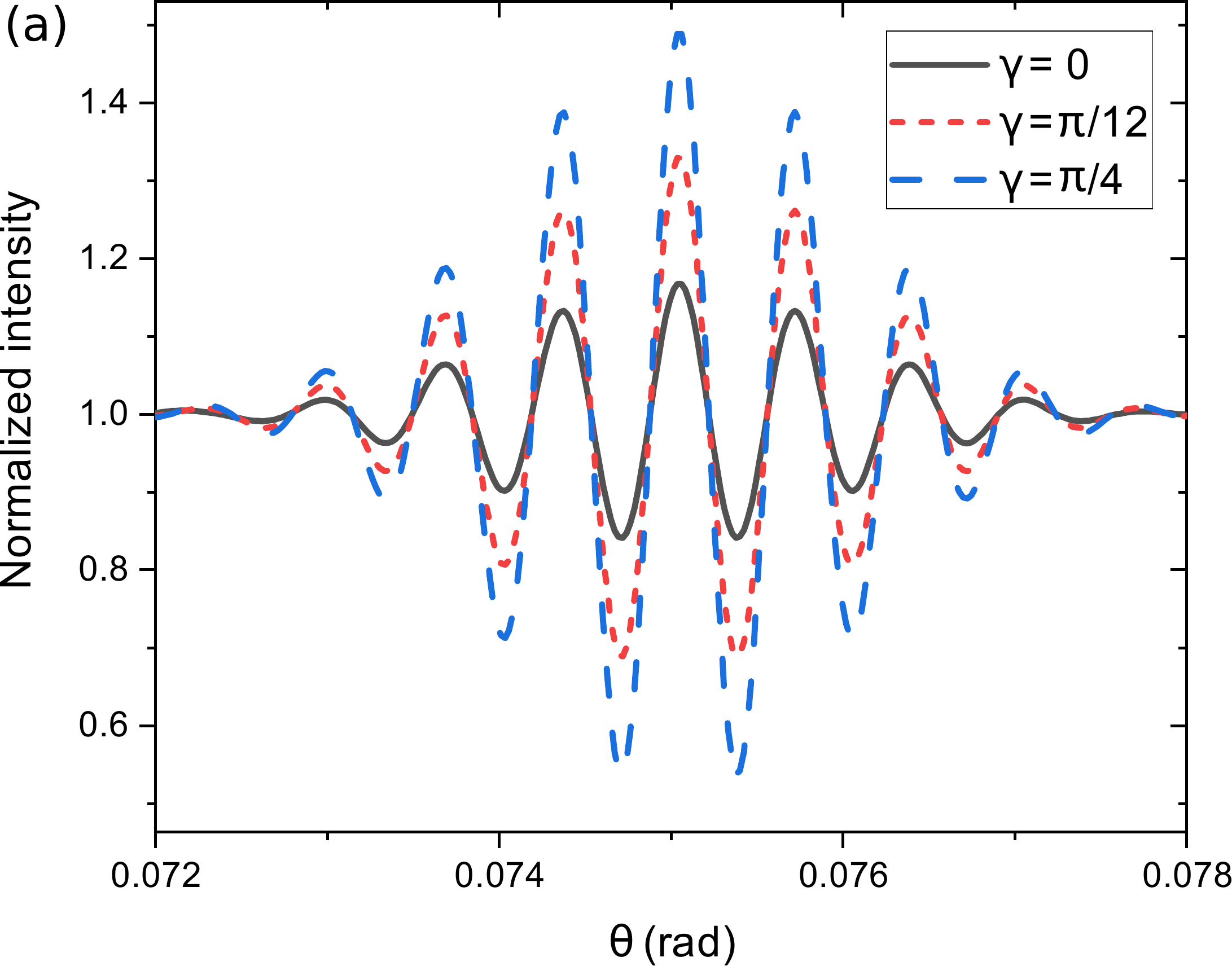} \\
\vspace{0.5cm}
\includegraphics[width=\columnwidth]{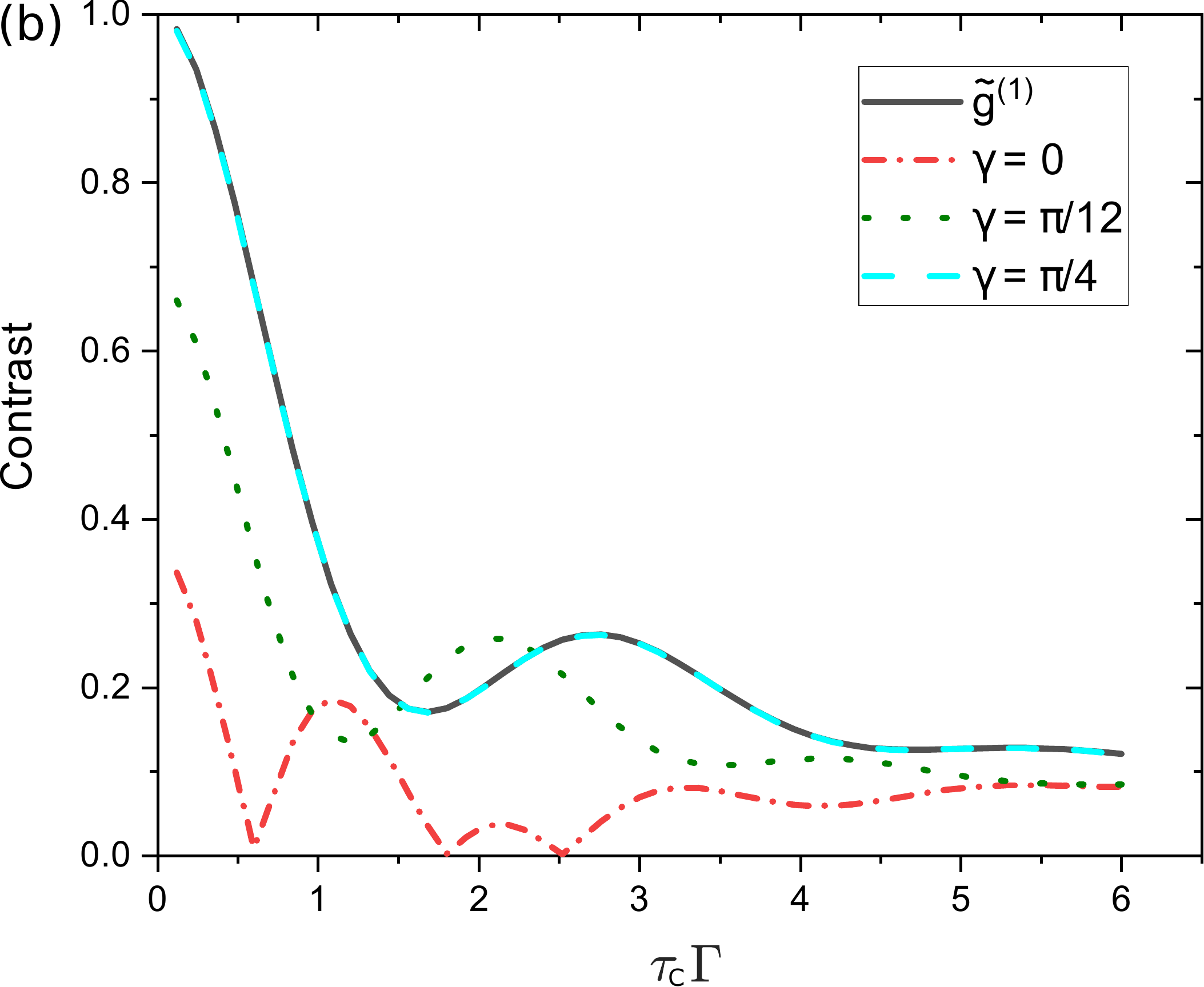}
\caption{(a)Angular fringes created by an extended cloud of randomly distributed scatterers, of transverse and longitudinal Gaussian widths $s_r = s_z = 500~\upmu$m, an incoming laser beam of plane wavefront with $\theta_0 = 4.3^{\circ}$, saturation parameter $s_0 = 5$, and $\tau_c \ll 1/(\sqrt{s_0} \, \Gamma)$, for different waveplate proper axis angles $\gamma$ with the incoming polarization. (b) Contrast at the center of the fringe pattern created by the extended cloud, for the same parameters as in (a), except for $\tau_c$ which is now varied from 0 to $6\Gamma$. We also plot for comparison $\tilde{g}^{(1)} (\tau_c)$ for a single atom, subject to a total saturation parameter $s = 2 s_0$ created by the superposition of the incoming and reflected excitation laser beams with perpendicular polarizations.}
\label{fig:fringes_cloud}
\end{figure}

The integral of Eq.(\ref{eq:Int_I}) has to be calculated numerically. We end up with angular fringes, as found in\,\cite{Piovella16}, and as plotted in\,Fig.\,\ref{fig:fringes_cloud}a for $s_r = s_z = 500~\upmu$m and $\theta_0 = 4.3^{\circ}$, and for different positions of the waveplate. The incoming laser field is a plane wave of saturation parameter $s_0 = 5$, and we choose a delay $\tau_c \ll 1/(\sqrt{s_0} \, \Gamma)$. Compared to the single-atom case, the fringes now present a finite angular envelope. We find numerically that the angular profile of the fringes depend on the laser excitation parameters and on the waveplate position only through its contrast, having otherwise a shape that depends only on the geometrical parameters of the system. This shape can be obtained analytically for $s_0 \ll 1\,$\cite{Moriya16}, which allows us to write:
\begin{eqnarray}
I(\mathbf{k})  & \propto & \bigg[1+ C (\tau_c,\Omega_l,\gamma)  \nonumber\\
&&  \e^{-2 \left(\theta_0 k s_z\right)^2 \left(\theta - \theta_0\right)^2} \cos \left(2 k h \theta_0 (\theta - \theta_0)\right) \bigg], 
\end{eqnarray}
with $C (\tau_c,\Omega_l,\gamma)$ the fringes contrast, computed in the general case through the numerical integration of Eq.\,(\ref{eq:Int_I}). We see that the fringe envelope is Gaussian, with a rms half-width of $s_{\theta} = 1/\left(2\theta_0 k s_z\right)$, while the spatial period of the fringes is given by $\Theta = \pi/(\theta_0 k h)$. 

For the case of parallel polarizations, the total intensity of the cloud is denoted as $I_{\parallel}$ and is given by\,\cite{Piovella16}:
\begin{eqnarray}
&&I_{\parallel}(\mathbf{k})  = \frac{N I_a}{\sqrt{2\pi} s_z}  \nonumber\\
&& \int \deriv z \, \e^{- \frac{(z+h)^2}{ 2 s_z^2}} \, \frac{s(z)}{1 + s(z)} \, \left[1 + \tilde{g}^{(1)}_z (\tau_c) \cos \left(2 k z \cos \theta \right) \right] . \nonumber\\
\label{eq:Int_I_parallel}
\end{eqnarray}
The contrast of the fringes $C_{\parallel} (\tau_c,\Omega_l) \equiv C (\tau_c,\Omega_l,\gamma = 0)$ is found by numerically integrating the above equation, and is shown in red dot-dashed line in Fig. \ref{fig:fringes_cloud}b. As for all values of $\gamma$ except $\gamma = \pi/4$ (see discussion below), this contrast is a complicated convolution of all correlation functions $\tilde{g}^{(1)}_z (\tau_c)$ for each position $z$ on the standing wave made by the interference of the incoming and reflected laser fields, and has no analytical expression to our knowledge. As another example, we also plot in Fig.\,\ref{fig:fringes_cloud}b the case of $\gamma = \pi/12$ in green dotted line, for which amplitude and polarization modulations must be taken into account.

For the case of mutually orthogonal polarization, $\gamma = \pi/4$, the total intensity of the cloud is denoted as $I_{\perp}$ and is equal to (see Appendix \ref{appendix1} for the calculations):
\begin{eqnarray}
I_{\perp}(\mathbf{k}) &=& N I_a \frac{s}{1+s}\bigg[1+ \tilde{g}^{(1)} (\tau_c) \nonumber\\
&&\int \deriv z \, \frac{\e^{- \frac{(z+h)^2}{2 s_z^2}}}{\sqrt{2\pi} s_z} \cos \left(2 k z \cos \theta_0 \right) \, \cos \left(2 k z \cos \theta \right) \bigg]  \nonumber\\
\label{eq:Int_I_perp}\\
&& = N I_a \frac{s}{1 + s} \bigg[1+ \tilde{g}^{(1)} (\tau_c)  \nonumber\\
&& \e^{-2 \left(\theta_0 k s_z\right)^2 \left(\theta - \theta_0\right)^2} \cos \left(2 k h \theta_0 (\theta - \theta_0)\right) \bigg]. \label{eq:perp_final}
\end{eqnarray}
For this particular case, the contrast at the center of the fringe pattern, around $\theta = \theta_0$, is simply given by:
\begin{equation}
C_{\perp} =  \tilde{g}^{(1)} (\tau_c) \ ,
\label{eq:contrast}
\end{equation}
thus directly equal to the first-order temporal correlation function of the light emitted by an atom subjected to a Rabi frequency $\Omega = \sqrt{2} d E_0/\hbar$, or equivalently, to a saturation parameter $s = 2 \Omega^2/\Gamma^2 = 4 d^2 E_0^2/(\hbar \Gamma)^2 = 2 s_0$.

%This intensity is plotted in Fig.\,\ref{fig:fringes_cloud}a, together with other intermediate positions of the half waveplate.
% We see that the geometric aspects of the fringes, their angular period and envelope, are identical, but their contrasts greatly vary for the different polarization configurations. The Gaussian envelope has an rms width of $s_{\theta} = 1/\left(2\theta_0 k s_z\right)$ and the spatial period is $\Theta = \pi/(\theta_0 k h)$.
 
The contrast obtained for perpendicular polarizations corresponds to the blue dashed line in Fig.\,\ref{fig:fringes_cloud}b, found by a numerical integration of Eq.\,(\ref{eq:Int_I}). As expected from Eq.\,(\ref{eq:contrast}), this curve is perfectly superimposed to the single atom case $\tilde{g}^{(1)} (\tau_c)$.  In this configuration, all atoms are subject to the same total saturation parameter $s = 2 s_0$, created by the superposition of the incoming and reflected excitation laser beams with mutual orthogonal polarizations.

%%%%%%%%%%%%%%%%%%%%%%%%%
%Conclusion
%%%%%%%%%%%%%%%%%%%%%%%%%

\section{Discussion and conclusion} \label{Sec:Conclusion}

In conclusion, we have shown that the MBS interferometer setup in the saturated regime has a contrast that depends on the linear polarization of the light reflected back to the atoms direction, when compared to the polarization of the light impinging on the mirror. The amplitude of the total electric field of the scattered light is the coherent superposition of four probability amplitudes, corresponding to four different scattering paths as shown in Fig. \ref{fig:Mach-Zehnder}a. Thanks to that, the fringes contrast survives for any polarization rotation of the reflected light, even in the case of mutual orthogonal polarizations. In this last case, the setup allows one to measure the first order temporal autocorrelation function $\tilde{g}^{(1)}(\tau)$ of the light scattered by the scatterers in the rotating referential frame of the incoming laser light. 

The feasibility of this measurement has been shown in a previous implementation of the interferometric setup \cite{Moriya16}; at that point, the saturation parameter and the distance to the real mirror weren't enough for probing the effects of the saturated spectrum of the atoms, and no waveplate was implemented. The main experimental constraints to be respected for the MBS signal to appear are the limits on the density and optical density of the sample in the detection direction, as discussed in Sec.\,\ref{Sec:LargeCloud}. Another important aspect is that the interferometer signal of eq. (\ref{eq:Int_I}) is composed solely of the scattered light, and the experimentalist must the able to separate it from the incoming light reflected close to the detection direction. The MBS interferometer signal appears for $\theta \sim \theta_0$, which configure a cone of directions making an angle $\sim \theta_0$ with respect to the normal of the mirror. At one point of this cone, we have the incoming reflected light, with wavevector $\mathbf{k}_0' = k\left(0,- \sin \theta_0, -\cos \theta_0 \right)$. So, the best direction for the detection is around the direction given by the wavevector $\mathbf{k} = k\left(0, \sin \theta_0, -\cos \theta_0 \right) = - \mathbf{k}_0$, the direction opposite to that of the incoming beam, which is separated from the direction of the reflected incoming beam by $2 \theta_0$ (see the experimental arrangement at \cite{Moriya16} for an implementation of this configuration). We also note that the MBS setup was recently applied to probe the coherence of the light scattered by a hot vapour in \cite{Cherroret2019robust}.

%A fair question to ask would be what is the advantage of using such a setup, when compared to other interferometric schemes that also allow to obtain directly $g^{(1)}(\tau)$? 
Although feasible and robust, the implementation and alignment of the MBS setup demand some work. It may thus not configure the best, or easiest, method to obtain $g^{(1)}(\tau)$ for any experimental configuration. However, it does present advantages that can be useful for specific experimental contexts. An advantage of the MBS technique over other interferometer setups also used to obtain $g^{(1)}(\tau)$, as for instance a MZ or Michelson interferometer, is the much less drastic requirement on the stability of the scanning mirror position: while one needs sub-wavelength steps and precision in a Michelson or MZ setup, the MBS setup requires steps in the order of a few hundredths of the smallest value of both parameters $c/\Omega_l$ and $c/\Gamma$, which is typically centimeter-sized for atoms, or hundreds of micrometers for quantum dots. On the other hand, it does not need fast photodetectors and electronics, as it is the case for measurements of $g^{(1)} (\tau)$ based on heterodyne techniques, such as the one implemented in Refs.\,\cite{Gutierrez19,Ferreira_2020} with correlations between time-resolved single photon detection events. When considered in a broader perspective, the MBS setup was already used in the past to evidence the coherence of the light emitted by the atoms in the saturated regime \cite{Moriya16}
 and by atoms in a hot vapour \cite{Cherroret2019robust}, where it allowed to identify a regime where the scattered light present coherences in spite of the Doppler broadening of the transition. The results of this article extend the capabilities of this setup, showing for instance that for the implementations above, simply adding a half waveplate will allow for the first-order correlations of light to be obtained without extra effort.

%This interferometric setup can be compared on one hand to other interferometers used to measure $g^{(1)}(\tau)$, as for example simple MZ or Michelson interferometers, where $g^{(1)}(\tau)$ can be obtained by monitoring the intensity of one of its two outputs, while the delay between both arms is scanned. An advantage of the MBS technique over these methods is the much less drastic requirement on the stability of the scanning mirror position: while one needs sub-wavelength steps and precision in a Michelson or MZ setup, the MBS setup requires steps in the order of a few hundredths of the smallest value of both parameters $c/\Omega_l$ and $c/\Gamma$, which is typically centimeter-sized for atoms, or hundreds of micrometers for quantum dots. On the other hand, the MBS setup can be compared to measurements of $g^{(1)} (\tau)$ based on heterodyne techniques, such as the one implemented in Refs.\,\cite{Gutierrez19,Ferreira_2020} with correlations between time-resolved single photon detection events. Our setup has the advantage of not needing fast photodetectors and electronics. To conclude, the MBS interferometric setup is an alternative method, which translates $g^{(1)}(\tau)$ into the contrast of angular interference fringes, that can be a more suitable approach for specific experimental constraints.

%%%%%%%%%%%%%%%%%%%%%%%%%
%Acknowledgements
%%%%%%%%%%%%%%%%%%%%%%%%%

\section*{Acknowledgements}

R. C. T. acknowledge funding from the Brazilian state agency FAPESP, grants no. 2018/23873-3 and 2013/07276-1. R. C. T., M.H and Ph. W. C. acknowledge funding from the French National Research Agency (projects QuaCor ANR19-CE47-0014-01). R. C. T. and M.H received support from the project CAPES-COFECUB (Ph879-17/CAPES 88887.130197/2017-01).

\begin{appendices}

\section{Integral of $I_{1,\perp}$}
\label{appendix1}

The integral of Eq.\,(\ref{eq:Int_I_perp}) is solved as follows:
\begin{widetext}
\begin{eqnarray}
&&\int_{- \infty}^{\infty} \deriv z \, \frac{\e^{- \frac{(z+h)^2}{2 s_z^2}}}{\sqrt{2\pi} s_z} \cos \left(2 k z \cos \theta_0 \right) \, \cos \left(2 k z \cos \theta \right) \\
&=& \int_{- \infty}^{\infty} \deriv z \, \frac{\e^{- \frac{(z+h)^2}{2 s_z^2}}}{ \sqrt{2\pi} s_z}\, \frac{\left(\e^{2i k z \cos \theta_0} + \e^{-2i k z \cos \theta_0}\right)}{2} \frac{\left(\e^{2i k z \cos \theta} + \e^{-2i k z \cos \theta}\right)}{2} \\
&=& \int_{- \infty}^{\infty} \deriv z \, \frac{\e^{- \frac{(z+h)^2}{2 s_z^2}}}{ 2\sqrt{2\pi} s_z}\, \left(\e^{2i k z (\cos \theta_0 + \cos\theta_0)} + \mathrm{c.c.} + \e^{2i k z (\cos \theta-\cos \theta_0)} + \mathrm{c.c.} \right) \\
&=& \, \left[ \frac{\e^{-2k^2 s_z^2 \left(\cos\theta + \cos \theta_0 \right)^2}}{2} \cos \left(2 k h \left(\cos \theta + \cos \theta_0 \right) \right) + \frac{\e^{-2k^2 s_z^2 \left(\cos\theta - \cos \theta_0 \right)^2}}{2} \cos \left(2 k h \left(\cos \theta - \cos \theta_0 \right) \right) \right] \ .
\label{eq:app1}
\end{eqnarray}
\end{widetext}
The argument of the first exponential of Eq.\,(\ref{eq:app1}) has a modulus much bigger than one for any value of $\theta$, given typical experimental values ($\theta_0 \gtrsim 1^{\circ}$, $s_z \gtrsim 1~$mm$~\gg \lambda$), and this exponential can be neglected to a very good approximation; moreover, the cosine that multiplies it vary too fast with $\theta$ for it to be detected, and it averages out to zero within the diffraction limit of the experimental detection setup. On the other hand, the second term induces an intensity modulation which can be experimentally detected for $\theta \sim \theta_0$. For $\theta, \theta_0 \ll 1$, $\theta \sim \theta_0$, we write $\cos\theta - \cos \theta_0 \simeq (\theta^2 - \theta_0^2)/2 \simeq \theta_0 (\theta - \theta_0)$. Replacing this in Eq.\,(\ref{eq:app1}), one gets:
\begin{multline}
\int_{- \infty}^{\infty} \deriv z \, \frac{\e^{- \frac{(z+h)^2}{2 s_z^2}}}{\sqrt{2\pi} s_z} \cos \left(2 k z \cos \theta_0 \right) \, \cos \left(2 k z \cos \theta \right) \\
\simeq \e^{-2 \left(\theta_0 k s_z\right)^2 \left(\theta - \theta_0\right)^2} \cos \left(2 k h \theta_0 (\theta - \theta_0)\right) \ .
\end{multline}

\end{appendices}

%%%%%%%%%%%%%%%%%%%%%%% References %%%%%%%%%%%%%%%%%%%%%%%%%

\bibliography{bibli}

\begin{thebibliography}{29}%
\makeatletter
\providecommand \@ifxundefined [1]{%
 \@ifx{#1\undefined}
}%
\providecommand \@ifnum [1]{%
 \ifnum #1\expandafter \@firstoftwo
 \else \expandafter \@secondoftwo
 \fi
}%
\providecommand \@ifx [1]{%
 \ifx #1\expandafter \@firstoftwo
 \else \expandafter \@secondoftwo
 \fi
}%
\providecommand \natexlab [1]{#1}%
\providecommand \enquote  [1]{``#1''}%
\providecommand \bibnamefont  [1]{#1}%
\providecommand \bibfnamefont [1]{#1}%
\providecommand \citenamefont [1]{#1}%
\providecommand \href@noop [0]{\@secondoftwo}%
\providecommand \href [0]{\begingroup \@sanitize@url \@href}%
\providecommand \@href[1]{\@@startlink{#1}\@@href}%
\providecommand \@@href[1]{\endgroup#1\@@endlink}%
\providecommand \@sanitize@url [0]{\catcode `\\12\catcode `\$12\catcode
  `\&12\catcode `\#12\catcode `\^12\catcode `\_12\catcode `\%12\relax}%
\providecommand \@@startlink[1]{}%
\providecommand \@@endlink[0]{}%
\providecommand \url  [0]{\begingroup\@sanitize@url \@url }%
\providecommand \@url [1]{\endgroup\@href {#1}{\urlprefix }}%
\providecommand \urlprefix  [0]{URL }%
\providecommand \Eprint [0]{\href }%
\providecommand \doibase [0]{https://doi.org/}%
\providecommand \selectlanguage [0]{\@gobble}%
\providecommand \bibinfo  [0]{\@secondoftwo}%
\providecommand \bibfield  [0]{\@secondoftwo}%
\providecommand \translation [1]{[#1]}%
\providecommand \BibitemOpen [0]{}%
\providecommand \bibitemStop [0]{}%
\providecommand \bibitemNoStop [0]{.\EOS\space}%
\providecommand \EOS [0]{\spacefactor3000\relax}%
\providecommand \BibitemShut  [1]{\csname bibitem#1\endcsname}%
\let\auto@bib@innerbib\@empty
%</preamble>
\bibitem [{\citenamefont {C.~Cohen-Tannoudji}\ and\ \citenamefont
  {Grynberg}(1998)}]{CCTAtom}%
  \BibitemOpen
  \bibfield  {author} {\bibinfo {author} {\bibfnamefont {J.~D.-R.}\
  \bibnamefont {C.~Cohen-Tannoudji}}\ and\ \bibinfo {author} {\bibfnamefont
  {G.}~\bibnamefont {Grynberg}},\ }\href@noop {} {\emph {\bibinfo {title}
  {Atom-photon interactions: basic processes and applications. Section V-D}}}\
  (\bibinfo  {publisher} {Wiley-VCH},\ \bibinfo {year} {1998})\BibitemShut
  {NoStop}%
\bibitem [{\citenamefont {Mollow}(1969)}]{Mollow69}%
  \BibitemOpen
  \bibfield  {author} {\bibinfo {author} {\bibfnamefont {B.~R.}\ \bibnamefont
  {Mollow}},\ }\bibfield  {title} {\bibinfo {title} {Power spectrum of light
  scattered by two-level systems},\ }\href
  {https://doi.org/10.1103/PhysRev.188.1969} {\bibfield  {journal} {\bibinfo
  {journal} {Phys. Rev.}\ }\textbf {\bibinfo {volume} {188}},\ \bibinfo {pages}
  {1969} (\bibinfo {year} {1969})}\BibitemShut {NoStop}%
\bibitem [{\citenamefont {Schrama}\ \emph {et~al.}(1992)\citenamefont
  {Schrama}, \citenamefont {Nienhuis}, \citenamefont {Dijkerman}, \citenamefont
  {Steijsiger},\ and\ \citenamefont {Heideman}}]{Schrama92}%
  \BibitemOpen
  \bibfield  {author} {\bibinfo {author} {\bibfnamefont {C.~A.}\ \bibnamefont
  {Schrama}}, \bibinfo {author} {\bibfnamefont {G.}~\bibnamefont {Nienhuis}},
  \bibinfo {author} {\bibfnamefont {H.~A.}\ \bibnamefont {Dijkerman}}, \bibinfo
  {author} {\bibfnamefont {C.}~\bibnamefont {Steijsiger}},\ and\ \bibinfo
  {author} {\bibfnamefont {H.~G.~M.}\ \bibnamefont {Heideman}},\ }\bibfield
  {title} {\bibinfo {title} {Intensity correlations between the components of
  the resonance fluorescence triplet},\ }\href
  {https://doi.org/10.1103/PhysRevA.45.8045} {\bibfield  {journal} {\bibinfo
  {journal} {Phys. Rev. A}\ }\textbf {\bibinfo {volume} {45}},\ \bibinfo
  {pages} {8045} (\bibinfo {year} {1992})}\BibitemShut {NoStop}%
\bibitem [{\citenamefont {Peiris}\ \emph {et~al.}(2015)\citenamefont {Peiris},
  \citenamefont {Petrak}, \citenamefont {Konthasinghe}, \citenamefont {Yu},
  \citenamefont {Niu},\ and\ \citenamefont {Muller}}]{Peiris15}%
  \BibitemOpen
  \bibfield  {author} {\bibinfo {author} {\bibfnamefont {M.}~\bibnamefont
  {Peiris}}, \bibinfo {author} {\bibfnamefont {B.}~\bibnamefont {Petrak}},
  \bibinfo {author} {\bibfnamefont {K.}~\bibnamefont {Konthasinghe}}, \bibinfo
  {author} {\bibfnamefont {Y.}~\bibnamefont {Yu}}, \bibinfo {author}
  {\bibfnamefont {Z.~C.}\ \bibnamefont {Niu}},\ and\ \bibinfo {author}
  {\bibfnamefont {A.}~\bibnamefont {Muller}},\ }\bibfield  {title} {\bibinfo
  {title} {Two-color photon correlations of the light scattered by a quantum
  dot},\ }\href {https://doi.org/10.1103/PhysRevB.91.195125} {\bibfield
  {journal} {\bibinfo  {journal} {Phys. Rev. B}\ }\textbf {\bibinfo {volume}
  {91}},\ \bibinfo {pages} {195125} (\bibinfo {year} {2015})}\BibitemShut
  {NoStop}%
\bibitem [{\citenamefont {L\'opez Carre\~no}\ \emph {et~al.}(2017)\citenamefont
  {L\'opez Carre\~no}, \citenamefont {del Valle},\ and\ \citenamefont
  {Laussy}}]{Carreno17}%
  \BibitemOpen
  \bibfield  {author} {\bibinfo {author} {\bibfnamefont {J.}~\bibnamefont
  {L\'opez Carre\~no}}, \bibinfo {author} {\bibfnamefont {E.}~\bibnamefont {del
  Valle}},\ and\ \bibinfo {author} {\bibfnamefont {F.}~\bibnamefont {Laussy}},\
  }\bibfield  {title} {\bibinfo {title} {Photon correlations from the mollow
  triplet},\ }\href {https://doi.org/10.1002/lpor.201700090} {\bibfield
  {journal} {\bibinfo  {journal} {Laser \& Photonics Reviews}\ }\textbf
  {\bibinfo {volume} {11}},\ \bibinfo {pages} {1700090} (\bibinfo {year}
  {2017})}\BibitemShut {NoStop}%
\bibitem [{\citenamefont {Aspect}\ \emph {et~al.}(1980)\citenamefont {Aspect},
  \citenamefont {Roger}, \citenamefont {Reynaud}, \citenamefont {Dalibard},\
  and\ \citenamefont {Cohen-Tannoudji}}]{Aspect80}%
  \BibitemOpen
  \bibfield  {author} {\bibinfo {author} {\bibfnamefont {A.}~\bibnamefont
  {Aspect}}, \bibinfo {author} {\bibfnamefont {G.}~\bibnamefont {Roger}},
  \bibinfo {author} {\bibfnamefont {S.}~\bibnamefont {Reynaud}}, \bibinfo
  {author} {\bibfnamefont {J.}~\bibnamefont {Dalibard}},\ and\ \bibinfo
  {author} {\bibfnamefont {C.}~\bibnamefont {Cohen-Tannoudji}},\ }\bibfield
  {title} {\bibinfo {title} {Time correlations between the two sidebands of the
  resonance fluorescence triplet},\ }\href
  {https://doi.org/10.1103/PhysRevLett.45.617} {\bibfield  {journal} {\bibinfo
  {journal} {Phys. Rev. Lett.}\ }\textbf {\bibinfo {volume} {45}},\ \bibinfo
  {pages} {617} (\bibinfo {year} {1980})}\BibitemShut {NoStop}%
\bibitem [{\citenamefont {Ulhaq}\ \emph {et~al.}(2012)\citenamefont {Ulhaq},
  \citenamefont {Weiler},\ and\ \citenamefont {Ulrich}}]{Ulhaq12}%
  \BibitemOpen
  \bibfield  {author} {\bibinfo {author} {\bibfnamefont {A.}~\bibnamefont
  {Ulhaq}}, \bibinfo {author} {\bibfnamefont {S.}~\bibnamefont {Weiler}},\ and\
  \bibinfo {author} {\bibfnamefont {S.~e.~a.}\ \bibnamefont {Ulrich}},\
  }\bibfield  {title} {\bibinfo {title} {Cascaded single-photon emission from
  the mollow triplet sidebands of a quantum dot},\ }\href
  {https://doi.org/10.1038/nphoton.2012.23} {\bibfield  {journal} {\bibinfo
  {journal} {Nature Photonics}\ }\textbf {\bibinfo {volume} {6}},\ \bibinfo
  {pages} {238} (\bibinfo {year} {2012})}\BibitemShut {NoStop}%
\bibitem [{\citenamefont {Portalupi}\ \emph {et~al.}(2016)\citenamefont
  {Portalupi}, \citenamefont {Widmann}, \citenamefont {Nawrath}, \citenamefont
  {Jetter}, \citenamefont {Michler}, \citenamefont {Wrachtrup},\ and\
  \citenamefont {Gerhardt}}]{Portalupi2016}%
  \BibitemOpen
  \bibfield  {author} {\bibinfo {author} {\bibfnamefont {S.~L.}\ \bibnamefont
  {Portalupi}}, \bibinfo {author} {\bibfnamefont {M.}~\bibnamefont {Widmann}},
  \bibinfo {author} {\bibfnamefont {C.}~\bibnamefont {Nawrath}}, \bibinfo
  {author} {\bibfnamefont {M.}~\bibnamefont {Jetter}}, \bibinfo {author}
  {\bibfnamefont {P.}~\bibnamefont {Michler}}, \bibinfo {author} {\bibfnamefont
  {J.}~\bibnamefont {Wrachtrup}},\ and\ \bibinfo {author} {\bibfnamefont
  {I.}~\bibnamefont {Gerhardt}},\ }\bibfield  {title} {\bibinfo {title}
  {Simultaneous faraday filtering of the mollow triplet sidebands with the
  cs-d1 clock transition},\ }\href {https://doi.org/10.1038/ncomms13632}
  {\bibfield  {journal} {\bibinfo  {journal} {Nature Communications}\ }\textbf
  {\bibinfo {volume} {7}},\ \bibinfo {pages} {13632} (\bibinfo {year}
  {2016})}\BibitemShut {NoStop}%
\bibitem [{\citenamefont {F~Schuda}\ and\ \citenamefont
  {Hercher}(1974)}]{Schuda74}%
  \BibitemOpen
  \bibfield  {author} {\bibinfo {author} {\bibfnamefont {C.~R. S.~J.}\
  \bibnamefont {F~Schuda}}\ and\ \bibinfo {author} {\bibfnamefont
  {M.}~\bibnamefont {Hercher}},\ }\href
  {https://doi.org/10.1088/0022-3700/7/7/002} {\bibfield  {journal} {\bibinfo
  {journal} {J. Phys. B: At. Mol. Phys.}\ }\textbf {\bibinfo {volume} {7}},\
  \bibinfo {pages} {L198} (\bibinfo {year} {1974})}\BibitemShut {NoStop}%
\bibitem [{\citenamefont {F~Y~Wu}\ and\ \citenamefont {Ezekiel}(1975)}]{Wu75}%
  \BibitemOpen
  \bibfield  {author} {\bibinfo {author} {\bibfnamefont {R.~E.~G.}\
  \bibnamefont {F~Y~Wu}}\ and\ \bibinfo {author} {\bibfnamefont
  {S.}~\bibnamefont {Ezekiel}},\ }\href
  {https://doi.org/10.1103/PhysRevLett.35.1426} {\bibfield  {journal} {\bibinfo
   {journal} {Phys. Rev. Lett.}\ }\textbf {\bibinfo {volume} {35}},\ \bibinfo
  {pages} {1426–} (\bibinfo {year} {1975})}\BibitemShut {NoStop}%
\bibitem [{\citenamefont {R~E~Grove}\ and\ \citenamefont
  {Ezekiel}(1977)}]{Grove77}%
  \BibitemOpen
  \bibfield  {author} {\bibinfo {author} {\bibfnamefont {F.~Y.~W.}\
  \bibnamefont {R~E~Grove}}\ and\ \bibinfo {author} {\bibfnamefont
  {S.}~\bibnamefont {Ezekiel}},\ }\href
  {https://doi.org/10.1103/PhysRevA.15.227} {\bibfield  {journal} {\bibinfo
  {journal} {Phys. Rev. A}\ }\textbf {\bibinfo {volume} {15}},\ \bibinfo
  {pages} {227–} (\bibinfo {year} {1977})}\BibitemShut {NoStop}%
\bibitem [{\citenamefont {W~Hartig}\ and\ \citenamefont
  {Walther}(1976)}]{Walther76}%
  \BibitemOpen
  \bibfield  {author} {\bibinfo {author} {\bibfnamefont {R.~S.}\ \bibnamefont
  {W~Hartig}, \bibfnamefont {W~Rasmussen}}\ and\ \bibinfo {author}
  {\bibfnamefont {H.}~\bibnamefont {Walther}},\ }\href
  {https://doi.org/10.1007/BF01409169} {\bibfield  {journal} {\bibinfo
  {journal} {Z. Phys. A}\ }\textbf {\bibinfo {volume} {278}},\ \bibinfo {pages}
  {205–} (\bibinfo {year} {1976})}\BibitemShut {NoStop}%
\bibitem [{\citenamefont {Stalgies}\ \emph {et~al.}(1996)\citenamefont
  {Stalgies}, \citenamefont {Siemers}, \citenamefont {Appasamy}, \citenamefont
  {Altevogt},\ and\ \citenamefont {Toschek}}]{Stalgies96}%
  \BibitemOpen
  \bibfield  {author} {\bibinfo {author} {\bibfnamefont {Y.}~\bibnamefont
  {Stalgies}}, \bibinfo {author} {\bibfnamefont {I.}~\bibnamefont {Siemers}},
  \bibinfo {author} {\bibfnamefont {B.}~\bibnamefont {Appasamy}}, \bibinfo
  {author} {\bibfnamefont {T.}~\bibnamefont {Altevogt}},\ and\ \bibinfo
  {author} {\bibfnamefont {P.~E.}\ \bibnamefont {Toschek}},\ }\bibfield
  {title} {\bibinfo {title} {The spectrum of single-atom resonance
  fluorescence},\ }\href {https://doi.org/10.1209/epl/i1996-00563-6} {\bibfield
   {journal} {\bibinfo  {journal} {Europhysics Letters ({EPL})}\ }\textbf
  {\bibinfo {volume} {35}},\ \bibinfo {pages} {259} (\bibinfo {year}
  {1996})}\BibitemShut {NoStop}%
\bibitem [{\citenamefont {G~Wrigge}(2008)}]{Wrigge08}%
  \BibitemOpen
  \bibfield  {author} {\bibinfo {author} {\bibfnamefont {J.~H. e.~a.}\
  \bibnamefont {G~Wrigge}, \bibfnamefont {I~Gerhardt}},\ }\bibfield  {title}
  {\bibinfo {title} {Efficient coupling of photons to a single molecule and the
  observation of its resonance fluorescence},\ }\href
  {https://doi.org/10.1038/nphys812} {\bibfield  {journal} {\bibinfo  {journal}
  {Nature Physics}\ }\textbf {\bibinfo {volume} {4}},\ \bibinfo {pages} {60}
  (\bibinfo {year} {2008})}\BibitemShut {NoStop}%
\bibitem [{\citenamefont {Flagg}\ \emph {et~al.}(2009)\citenamefont {Flagg},
  \citenamefont {Muller}, \citenamefont {Robertson}, \citenamefont {Founta},
  \citenamefont {Deppe}, \citenamefont {Xiao}, \citenamefont {Ma},
  \citenamefont {Salamo},\ and\ \citenamefont {Shih}}]{Flagg09}%
  \BibitemOpen
  \bibfield  {author} {\bibinfo {author} {\bibfnamefont {E.~B.}\ \bibnamefont
  {Flagg}}, \bibinfo {author} {\bibfnamefont {A.}~\bibnamefont {Muller}},
  \bibinfo {author} {\bibfnamefont {J.~W.}\ \bibnamefont {Robertson}}, \bibinfo
  {author} {\bibfnamefont {S.}~\bibnamefont {Founta}}, \bibinfo {author}
  {\bibfnamefont {D.~G.}\ \bibnamefont {Deppe}}, \bibinfo {author}
  {\bibfnamefont {M.}~\bibnamefont {Xiao}}, \bibinfo {author} {\bibfnamefont
  {W.}~\bibnamefont {Ma}}, \bibinfo {author} {\bibfnamefont {G.~J.}\
  \bibnamefont {Salamo}},\ and\ \bibinfo {author} {\bibfnamefont {C.~K.}\
  \bibnamefont {Shih}},\ }\bibfield  {title} {\bibinfo {title} {Resonantly
  driven coherent oscillations in a solid-state quantum emitter},\ }\href
  {https://doi.org/10.1038/nphys1184} {\bibfield  {journal} {\bibinfo
  {journal} {Nature Physics}\ }\textbf {\bibinfo {volume} {5}},\ \bibinfo
  {pages} {203} (\bibinfo {year} {2009})}\BibitemShut {NoStop}%
\bibitem [{\citenamefont {Lagoudakis}\ \emph {et~al.}(2017)\citenamefont
  {Lagoudakis}, \citenamefont {Fischer}, \citenamefont {Sarmiento},
  \citenamefont {McMahon}, \citenamefont {Radulaski}, \citenamefont {Zhang},
  \citenamefont {Kelaita}, \citenamefont {Dory}, \citenamefont {M\"uller},\
  and\ \citenamefont {Vu\ifmmode \check{c}\else
  \v{c}\fi{}kovi\ifmmode~\acute{c}\else \'{c}\fi{}}}]{Lagoudakis2017}%
  \BibitemOpen
  \bibfield  {author} {\bibinfo {author} {\bibfnamefont {K.~G.}\ \bibnamefont
  {Lagoudakis}}, \bibinfo {author} {\bibfnamefont {K.~A.}\ \bibnamefont
  {Fischer}}, \bibinfo {author} {\bibfnamefont {T.}~\bibnamefont {Sarmiento}},
  \bibinfo {author} {\bibfnamefont {P.~L.}\ \bibnamefont {McMahon}}, \bibinfo
  {author} {\bibfnamefont {M.}~\bibnamefont {Radulaski}}, \bibinfo {author}
  {\bibfnamefont {J.~L.}\ \bibnamefont {Zhang}}, \bibinfo {author}
  {\bibfnamefont {Y.}~\bibnamefont {Kelaita}}, \bibinfo {author} {\bibfnamefont
  {C.}~\bibnamefont {Dory}}, \bibinfo {author} {\bibfnamefont {K.}~\bibnamefont
  {M\"uller}},\ and\ \bibinfo {author} {\bibfnamefont {J.}~\bibnamefont
  {Vu\ifmmode \check{c}\else \v{c}\fi{}kovi\ifmmode~\acute{c}\else
  \'{c}\fi{}}},\ }\bibfield  {title} {\bibinfo {title} {Observation of mollow
  triplets with tunable interactions in double lambda systems of individual
  hole spins},\ }\href {https://doi.org/10.1103/PhysRevLett.118.013602}
  {\bibfield  {journal} {\bibinfo  {journal} {Phys. Rev. Lett.}\ }\textbf
  {\bibinfo {volume} {118}},\ \bibinfo {pages} {013602} (\bibinfo {year}
  {2017})}\BibitemShut {NoStop}%
\bibitem [{\citenamefont {Loudon}(2000)}]{Loudon}%
  \BibitemOpen
  \bibfield  {author} {\bibinfo {author} {\bibfnamefont {R.}~\bibnamefont
  {Loudon}},\ }\href@noop {} {\emph {\bibinfo {title} {The quantum theory of
  light}}}\ (\bibinfo  {publisher} {OUP Oxford},\ \bibinfo {year}
  {2000})\BibitemShut {NoStop}%
\bibitem [{\citenamefont {Okoshi}\ \emph {et~al.}(1980)\citenamefont {Okoshi},
  \citenamefont {Kikuchi},\ and\ \citenamefont {Nakayama}}]{Okoshi_1980}%
  \BibitemOpen
  \bibfield  {author} {\bibinfo {author} {\bibfnamefont {T.}~\bibnamefont
  {Okoshi}}, \bibinfo {author} {\bibfnamefont {K.}~\bibnamefont {Kikuchi}},\
  and\ \bibinfo {author} {\bibfnamefont {A.}~\bibnamefont {Nakayama}},\
  }\bibfield  {title} {\bibinfo {title} {Novel method for high resolution
  measurement of laser output spectrum},\ }\href@noop {} {\bibfield  {journal}
  {\bibinfo  {journal} {Electron. Lett.}\ }\textbf {\bibinfo {volume} {16}},\
  \bibinfo {pages} {630} (\bibinfo {year} {1980})}\BibitemShut {NoStop}%
\bibitem [{\citenamefont {Ferreira}\ \emph {et~al.}(2020)\citenamefont
  {Ferreira}, \citenamefont {Bachelard}, \citenamefont {Guerin}, \citenamefont
  {Kaiser},\ and\ \citenamefont {Fouché}}]{Ferreira_2020}%
  \BibitemOpen
  \bibfield  {author} {\bibinfo {author} {\bibfnamefont {D.}~\bibnamefont
  {Ferreira}}, \bibinfo {author} {\bibfnamefont {R.}~\bibnamefont {Bachelard}},
  \bibinfo {author} {\bibfnamefont {W.}~\bibnamefont {Guerin}}, \bibinfo
  {author} {\bibfnamefont {R.}~\bibnamefont {Kaiser}},\ and\ \bibinfo {author}
  {\bibfnamefont {M.}~\bibnamefont {Fouché}},\ }\bibfield  {title} {\bibinfo
  {title} {Connecting field and intensity correlations: The siegert relation
  and how to test it},\ }\href {https://doi.org/10.1119/10.0001630} {\bibfield
  {journal} {\bibinfo  {journal} {American Journal of Physics}\ }\textbf
  {\bibinfo {volume} {88}},\ \bibinfo {pages} {831} (\bibinfo {year} {2020})},\
  \Eprint {https://arxiv.org/abs/https://doi.org/10.1119/10.0001630}
  {https://doi.org/10.1119/10.0001630} \BibitemShut {NoStop}%
\bibitem [{\citenamefont {Nakayama}\ \emph {et~al.}(2010)\citenamefont
  {Nakayama}, \citenamefont {Yoshikawa}, \citenamefont {Matsumoto},
  \citenamefont {Torii},\ and\ \citenamefont {Kuga}}]{Nakayama10}%
  \BibitemOpen
  \bibfield  {author} {\bibinfo {author} {\bibfnamefont {K.}~\bibnamefont
  {Nakayama}}, \bibinfo {author} {\bibfnamefont {Y.}~\bibnamefont {Yoshikawa}},
  \bibinfo {author} {\bibfnamefont {H.}~\bibnamefont {Matsumoto}}, \bibinfo
  {author} {\bibfnamefont {Y.}~\bibnamefont {Torii}},\ and\ \bibinfo {author}
  {\bibfnamefont {T.}~\bibnamefont {Kuga}},\ }\bibfield  {title} {\bibinfo
  {title} {Precise intensity correlation measurement for atomic resonance
  fluorescence from optical molasses},\ }\href
  {https://doi.org/10.1364/OE.18.006604} {\bibfield  {journal} {\bibinfo
  {journal} {Opt. Express}\ }\textbf {\bibinfo {volume} {18}},\ \bibinfo
  {pages} {6604} (\bibinfo {year} {2010})}\BibitemShut {NoStop}%
\bibitem [{\citenamefont {Ortiz-Guti\'errez}\ \emph {et~al.}(2019)\citenamefont
  {Ortiz-Guti\'errez}, \citenamefont {Teixeira}, \citenamefont {Eloy},
  \citenamefont {da~Silva}, \citenamefont {Kaiser}, \citenamefont {Bachelard},\
  and\ \citenamefont {Fouch\'e}}]{Gutierrez19}%
  \BibitemOpen
  \bibfield  {author} {\bibinfo {author} {\bibfnamefont {L.}~\bibnamefont
  {Ortiz-Guti\'errez}}, \bibinfo {author} {\bibfnamefont {R.}~\bibnamefont
  {Teixeira}}, \bibinfo {author} {\bibfnamefont {A.}~\bibnamefont {Eloy}},
  \bibinfo {author} {\bibfnamefont {D.}~\bibnamefont {da~Silva}}, \bibinfo
  {author} {\bibfnamefont {R.}~\bibnamefont {Kaiser}}, \bibinfo {author}
  {\bibfnamefont {R.}~\bibnamefont {Bachelard}},\ and\ \bibinfo {author}
  {\bibfnamefont {M.}~\bibnamefont {Fouch\'e}},\ }\bibfield  {title} {\bibinfo
  {title} {Mollow triplet in cold atoms},\ }\href
  {https://doi.org/10.1088/1367-2630/ab3ca9} {\bibfield  {journal} {\bibinfo
  {journal} {New Journal of Physics}\ }\textbf {\bibinfo {volume} {21}},\
  \bibinfo {pages} {093019} (\bibinfo {year} {2019})}\BibitemShut {NoStop}%
\bibitem [{\citenamefont {Greffet}(1991)}]{Greffet91}%
  \BibitemOpen
  \bibfield  {author} {\bibinfo {author} {\bibfnamefont {J.~J.}\ \bibnamefont
  {Greffet}},\ }\bibfield  {title} {\bibinfo {title} {Backscattering of
  s-polarized light from a cloud of small particles above a dielectric
  substrate},\ }\href {https://doi.org/10.1088/0959-7174/1/3/006} {\bibfield
  {journal} {\bibinfo  {journal} {Waves in Random Media}\ }\textbf {\bibinfo
  {volume} {3}},\ \bibinfo {pages} {S65} (\bibinfo {year} {1991})}\BibitemShut
  {NoStop}%
\bibitem [{\citenamefont {Labeyrie}\ \emph {et~al.}(2000)\citenamefont
  {Labeyrie}, \citenamefont {M{\"u}ller}, \citenamefont {Wiersma},
  \citenamefont {Miniatura},\ and\ \citenamefont
  {Kaiser}}]{Labeyrie2000observation}%
  \BibitemOpen
  \bibfield  {author} {\bibinfo {author} {\bibfnamefont {G.}~\bibnamefont
  {Labeyrie}}, \bibinfo {author} {\bibfnamefont {C.}~\bibnamefont
  {M{\"u}ller}}, \bibinfo {author} {\bibfnamefont {D.}~\bibnamefont {Wiersma}},
  \bibinfo {author} {\bibfnamefont {C.}~\bibnamefont {Miniatura}},\ and\
  \bibinfo {author} {\bibfnamefont {R.}~\bibnamefont {Kaiser}},\ }\bibfield
  {title} {\bibinfo {title} {Observation of coherent backscattering of light by
  cold atoms},\ }\href {https://doi.org/10.1088/1464-4266/2/5/316} {\bibfield
  {journal} {\bibinfo  {journal} {Journal of Optics B: Quantum and
  Semiclassical Optics}\ }\textbf {\bibinfo {volume} {2}},\ \bibinfo {pages}
  {672} (\bibinfo {year} {2000})}\BibitemShut {NoStop}%
\bibitem [{\citenamefont {Moriya}\ \emph {et~al.}(2016)\citenamefont {Moriya},
  \citenamefont {Shiozaki}, \citenamefont {Teixeira}, \citenamefont {Maximo},
  \citenamefont {Piovella}, \citenamefont {Bachelard}, \citenamefont {Kaiser},\
  and\ \citenamefont {Courteille}}]{Moriya16}%
  \BibitemOpen
  \bibfield  {author} {\bibinfo {author} {\bibfnamefont {P.~H.}\ \bibnamefont
  {Moriya}}, \bibinfo {author} {\bibfnamefont {R.~F.}\ \bibnamefont
  {Shiozaki}}, \bibinfo {author} {\bibfnamefont {R.~C.}\ \bibnamefont
  {Teixeira}}, \bibinfo {author} {\bibfnamefont {C.~E.}\ \bibnamefont
  {Maximo}}, \bibinfo {author} {\bibfnamefont {N.}~\bibnamefont {Piovella}},
  \bibinfo {author} {\bibfnamefont {R.}~\bibnamefont {Bachelard}}, \bibinfo
  {author} {\bibfnamefont {R.}~\bibnamefont {Kaiser}},\ and\ \bibinfo {author}
  {\bibfnamefont {P.~W.}\ \bibnamefont {Courteille}},\ }\bibfield  {title}
  {\bibinfo {title} {Coherent backscattering of inelastic photons from atoms
  and their mirror images},\ }\href
  {https://doi.org/10.1103/PhysRevA.94.053806} {\bibfield  {journal} {\bibinfo
  {journal} {Physical Review A}\ }\textbf {\bibinfo {volume} {94}},\ \bibinfo
  {pages} {053806} (\bibinfo {year} {2016})}\BibitemShut {NoStop}%
\bibitem [{\citenamefont {Piovella}\ \emph {et~al.}(2017)\citenamefont
  {Piovella}, \citenamefont {Teixeira}, \citenamefont {Kaiser}, \citenamefont
  {Courteille},\ and\ \citenamefont {Bachelard}}]{Piovella16}%
  \BibitemOpen
  \bibfield  {author} {\bibinfo {author} {\bibfnamefont {N.}~\bibnamefont
  {Piovella}}, \bibinfo {author} {\bibfnamefont {R.~C.}\ \bibnamefont
  {Teixeira}}, \bibinfo {author} {\bibfnamefont {R.}~\bibnamefont {Kaiser}},
  \bibinfo {author} {\bibfnamefont {P.~W.}\ \bibnamefont {Courteille}},\ and\
  \bibinfo {author} {\bibfnamefont {R.}~\bibnamefont {Bachelard}},\ }\bibfield
  {title} {\bibinfo {title} {Mirror-assisted coherent backscattering from the
  mollow sidebands},\ }\href {https://doi.org/10.1103/PhysRevA.96.053852}
  {\bibfield  {journal} {\bibinfo  {journal} {Physical Review A}\ }\textbf
  {\bibinfo {volume} {96}},\ \bibinfo {pages} {053852} (\bibinfo {year}
  {2017})}\BibitemShut {NoStop}%
\bibitem [{\citenamefont {Toyli}\ \emph {et~al.}(2016)\citenamefont {Toyli},
  \citenamefont {Eddins}, \citenamefont {Boutin}, \citenamefont {Puri},
  \citenamefont {Hover}, \citenamefont {Bolkhovsky}, \citenamefont {Oliver},
  \citenamefont {Blais},\ and\ \citenamefont {Siddiqi}}]{Toyli2016resonance}%
  \BibitemOpen
  \bibfield  {author} {\bibinfo {author} {\bibfnamefont {D.~M.}\ \bibnamefont
  {Toyli}}, \bibinfo {author} {\bibfnamefont {A.~W.}\ \bibnamefont {Eddins}},
  \bibinfo {author} {\bibfnamefont {S.}~\bibnamefont {Boutin}}, \bibinfo
  {author} {\bibfnamefont {S.}~\bibnamefont {Puri}}, \bibinfo {author}
  {\bibfnamefont {D.}~\bibnamefont {Hover}}, \bibinfo {author} {\bibfnamefont
  {V.}~\bibnamefont {Bolkhovsky}}, \bibinfo {author} {\bibfnamefont {W.~D.}\
  \bibnamefont {Oliver}}, \bibinfo {author} {\bibfnamefont {A.}~\bibnamefont
  {Blais}},\ and\ \bibinfo {author} {\bibfnamefont {I.}~\bibnamefont
  {Siddiqi}},\ }\bibfield  {title} {\bibinfo {title} {Resonance fluorescence
  from an artificial atom in squeezed vacuum},\ }\href
  {https://doi.org/10.1103/PhysRevX.6.031004} {\bibfield  {journal} {\bibinfo
  {journal} {Physical Review X}\ }\textbf {\bibinfo {volume} {6}},\ \bibinfo
  {pages} {031004} (\bibinfo {year} {2016})}\BibitemShut {NoStop}%
\bibitem [{\citenamefont {Keitel}\ \emph {et~al.}(1995)\citenamefont {Keitel},
  \citenamefont {Knight}, \citenamefont {Narducci},\ and\ \citenamefont
  {Scully}}]{Keitel1995resonance}%
  \BibitemOpen
  \bibfield  {author} {\bibinfo {author} {\bibfnamefont {C.~H.}\ \bibnamefont
  {Keitel}}, \bibinfo {author} {\bibfnamefont {P.~L.}\ \bibnamefont {Knight}},
  \bibinfo {author} {\bibfnamefont {L.~M.}\ \bibnamefont {Narducci}},\ and\
  \bibinfo {author} {\bibfnamefont {M.~O.}\ \bibnamefont {Scully}},\ }\bibfield
   {title} {\bibinfo {title} {Resonance fluorescence in a tailored vacuum},\
  }\href {https://doi.org/10.1016/0030-4018(95)00161-Z} {\bibfield  {journal}
  {\bibinfo  {journal} {Optics communications}\ }\textbf {\bibinfo {volume}
  {118}},\ \bibinfo {pages} {143} (\bibinfo {year} {1995})}\BibitemShut
  {NoStop}%
\bibitem [{\citenamefont {Cherroret}\ \emph {et~al.}(2019)\citenamefont
  {Cherroret}, \citenamefont {Hemmerling}, \citenamefont {Nador}, \citenamefont
  {Walraven},\ and\ \citenamefont {Kaiser}}]{Cherroret2019robust}%
  \BibitemOpen
  \bibfield  {author} {\bibinfo {author} {\bibfnamefont {N.}~\bibnamefont
  {Cherroret}}, \bibinfo {author} {\bibfnamefont {M.}~\bibnamefont
  {Hemmerling}}, \bibinfo {author} {\bibfnamefont {V.}~\bibnamefont {Nador}},
  \bibinfo {author} {\bibfnamefont {J.~T.~M.}\ \bibnamefont {Walraven}},\ and\
  \bibinfo {author} {\bibfnamefont {R.}~\bibnamefont {Kaiser}},\ }\bibfield
  {title} {\bibinfo {title} {Robust coherent transport of light in multilevel
  hot atomic vapors},\ }\href {https://doi.org/10.1103/PhysRevLett.122.183203}
  {\bibfield  {journal} {\bibinfo  {journal} {Physical review letters}\
  }\textbf {\bibinfo {volume} {122}},\ \bibinfo {pages} {183203} (\bibinfo
  {year} {2019})}\BibitemShut {NoStop}%
\bibitem [{\citenamefont {Bienaimé}\ \emph {et~al.}(2011)\citenamefont
  {Bienaimé}, \citenamefont {Petruzzo}, \citenamefont {Bigerni}, \citenamefont
  {Piovella},\ and\ \citenamefont {Kaiser}}]{Bienaime_2011}%
  \BibitemOpen
  \bibfield  {author} {\bibinfo {author} {\bibfnamefont {T.}~\bibnamefont
  {Bienaimé}}, \bibinfo {author} {\bibfnamefont {M.}~\bibnamefont {Petruzzo}},
  \bibinfo {author} {\bibfnamefont {D.}~\bibnamefont {Bigerni}}, \bibinfo
  {author} {\bibfnamefont {N.}~\bibnamefont {Piovella}},\ and\ \bibinfo
  {author} {\bibfnamefont {R.}~\bibnamefont {Kaiser}},\ }\bibfield  {title}
  {\bibinfo {title} {Atom and photon measurement in cooperative scattering by
  cold atoms},\ }\href {https://doi.org/10.1080/09500340.2011.594911}
  {\bibfield  {journal} {\bibinfo  {journal} {Journal of Modern Optics}\
  }\textbf {\bibinfo {volume} {58}},\ \bibinfo {pages} {1942} (\bibinfo {year}
  {2011})},\ \Eprint
  {https://arxiv.org/abs/https://doi.org/10.1080/09500340.2011.594911}
  {https://doi.org/10.1080/09500340.2011.594911} \BibitemShut {NoStop}%
\end{thebibliography}%

\end{document}